\documentclass[%
reprint,
superscriptaddress,
 amsmath,amssymb,
 aps,
prx
]{revtex4-2}

\usepackage{graphicx}% Include figure files
\usepackage{dcolumn}% Align table columns on decimal point
\usepackage{bm}% bold math
\usepackage[T1]{fontenc}

\usepackage[utf8]{inputenc}
\usepackage[version=4]{mhchem}
\usepackage{newtxtext}
\usepackage{newtxmath}
\usepackage{natbib}
\usepackage{hyperref}
\hypersetup{
    colorlinks = true,
    urlcolor   = blue,
    citecolor  = blue,
}
\usepackage[export]{adjustbox}
\usepackage{booktabs}

\newcommand{\dd}[0]{\mathrm{d}} % differential d
\newcommand{\pd}[0]{\partial} % partrial d
\newcommand{\peclet}[0]{\textrm{Pe}}
\newcommand{\sherwood}[0]{\textrm{Sh}}
\newcommand{\reynolds}[0]{\textrm{Re}}
\newcommand{\rparticle}[0]{\beta}
\newcommand{\modsherwood}[0]{\widetilde{\textrm{Sh}}}
\newcommand{\diffusiveflux}[0]{\Phi_{\textrm{D}}}
\newcommand{\cliftflux}[0]{\Phi_{\textrm{Cl}}}
\newcommand{\advectiveflux}[0]{\Phi_{\textrm{A}}}
\newcommand{\code}[1]{\texttt{#1}}

\newcommand{\Ito}[0]{Itô}

\begin{document}

\title{
Bridging advection and diffusion in the encounter dynamics of sedimenting marine snow
}

\author{Jan Turczynowicz}
\thanks{These authors contributed equally to this work.}
\affiliation{Faculty of Physics, University of Warsaw, Pasteura 5, 02-093 Warsaw, Poland}
\affiliation{Fenix Science Club, Aleja Stanów Zjednoczonych 24, 03-964 Warsaw, Poland.}

\author{Radost Waszkiewicz}
\thanks{These authors contributed equally to this work.}
\affiliation{Institute of Physics, Polish Academy of Sciences, Aleja Lotnikow 32/46, PL-02668 Warsaw, Poland}
\affiliation{Fenix Science Club, Aleja Stanów Zjednoczonych 24, 03-964 Warsaw, Poland.}

\author{Jonasz Słomka}
\thanks{Corresponding authors: jslomka@ethz.ch (JS); mklis@fuw.edu.pl (ML).}
\affiliation{Institute of Environmental Engineering, Department of Civil, Environmental and Geomatic Engineering, ETH Zurich, Zurich, Switzerland}

\author{Maciej Lisicki}
\thanks{Corresponding authors: jslomka@ethz.ch (JS); mklis@fuw.edu.pl (ML).}
\affiliation{Faculty of Physics, University of Warsaw, Pasteura 5, 02-093 Warsaw, Poland}

\date{\today}             

\begin{abstract}
Sinking marine snow particles, composed primarily of organic matter, control the global export of photosynthetically fixed carbon from the ocean surface to depth. 
The fate of sedimenting marine snow particles is in part regulated by their encounters with suspended, micron-sized objects, which leads to mass accretion by the particles and potentially alters their buoyancy, and with bacteria that can colonize the particles and degrade them. Their collision rates are typically calculated using two types of models focusing either on direct (ballistic) interception with a finite interaction range, or advective-diffusive capture with a zero interaction range. Yet, since the range of applicability of the two models is unknown, and many relevant marine encounter scenarios span across both regimes, quantifying such encounters remains challenging, mainly because the two models yield asymptotically different predictions at high P\'{e}clet numbers. Here, we reconcile the two limiting approaches by quantifying the encounters in the general case using a combination of theoretical analysis and numerical simulations. Specifically, by solving the advection-diffusion equation in Stokes flow around a sphere to model mass transfer to a large sinking particle by small yet finite-sized objects, we determine a new formula for the Sherwood number (dimensionless encounter rate) as a function of two dimensionless parameters: the P\'{e}clet number and the ratio of small to large particle sizes. Contrary to the common assumption, we find that diffusion can still play a significant role in generating encounters even at high P\'{e}clet numbers. 
We predict that in many scenarios, at P\'{e}clet numbers as high as $10^6$, the direct interception model underestimates the encounter rate by up to two orders of magnitude.
This overlooked contribution of diffusion to encounters suggests that important processes affecting the fate of marine snow, such as colonization by bacteria and plankton or changes in buoyancy induced by accretion of neutrally buoyant gels, may proceed at a rate much faster than previously thought.
\end{abstract}

\maketitle

\section{Introduction}

The oceans play a central role in capturing anthropogenic \ce{CO2}, primarily through dissolution processes, resulting in a significant portion of it being stored within seawater~\citep{Broecker_1982,Sabine_2004,Sarmiento_2006}. A fraction of the dissolved \ce{CO2} is transformed into organic compounds by the photosynthetic activity of phytoplankton~\citep{Guidi_2016} dwelling in the well-mixed euphotic zone, which extends from the surface to a depth of about 100 m~\citep{Buesseler_2009,Buesseler_2020}, and then is further converted into particulate matter known as marine snow. Marine snow particles form through aggregation of dead or senescent cells, detritus, organic and inorganic matter, and span many orders of magnitude in size and sinking speed~\citep{Duret_2018,Cael_2021,Clements_2022,Williams_2022}. Some of them sink beneath the mixing layer and start a journey to depth through mostly quiescent waters. This sedimentation-driven process (and, to a lesser extent, active transport by migrating organisms at intermediate depth~\citep{Boyd_2019}), called the biological carbon pump, is a significant mechanism of \ce{CO2} sequestration on the seabed~\citep{Boyd_2019,Sarmiento_2006}. Field observations, described by the Martin curve~\citep{Martin_1987}, show that as little as 10\% of the carbon sediment reaches depths beyond 200 m below the euphotic zone~\citep{Buesseler_2009}, and the carbon flux decays rapidly with depth~\citep{Olli_2015,Kenneth_2018,Buesseler_2007,Middelburg_2019,Jang_2024,Armstrong_2001,Giering_2017,Laufkötter_2017,Takeuchi_2024,Marsay_2015}. Quantification of this flux decay requires understanding of the underlying microscale interactions within sedimenting matter~\cite{Nguyen_2022}.

\begin{figure*}
    \centering
    \begin{tabular}{lr}        
         \includegraphics[height=0.36\linewidth,valign=m]{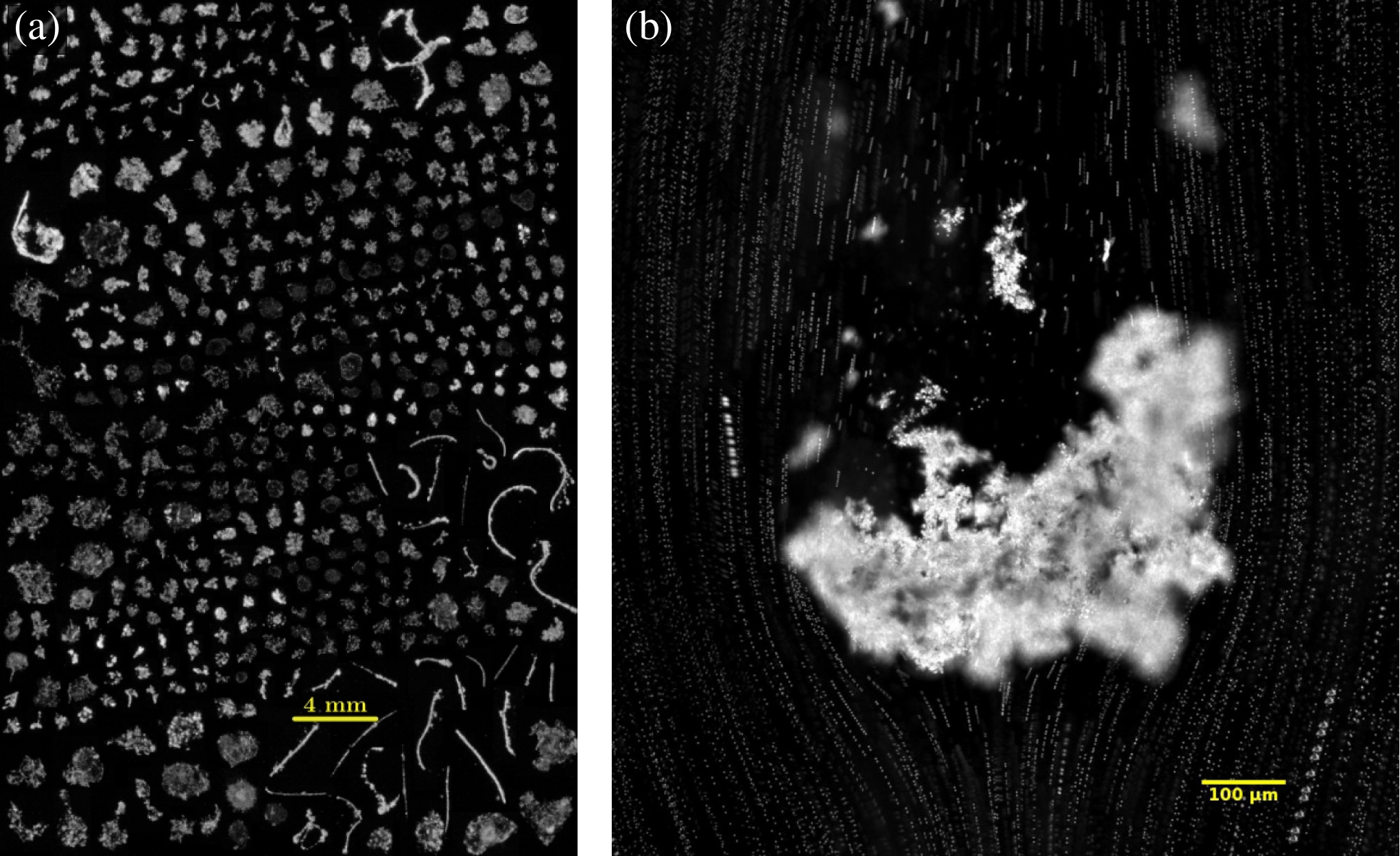} &
         \includegraphics[height=0.36\linewidth,valign=m]{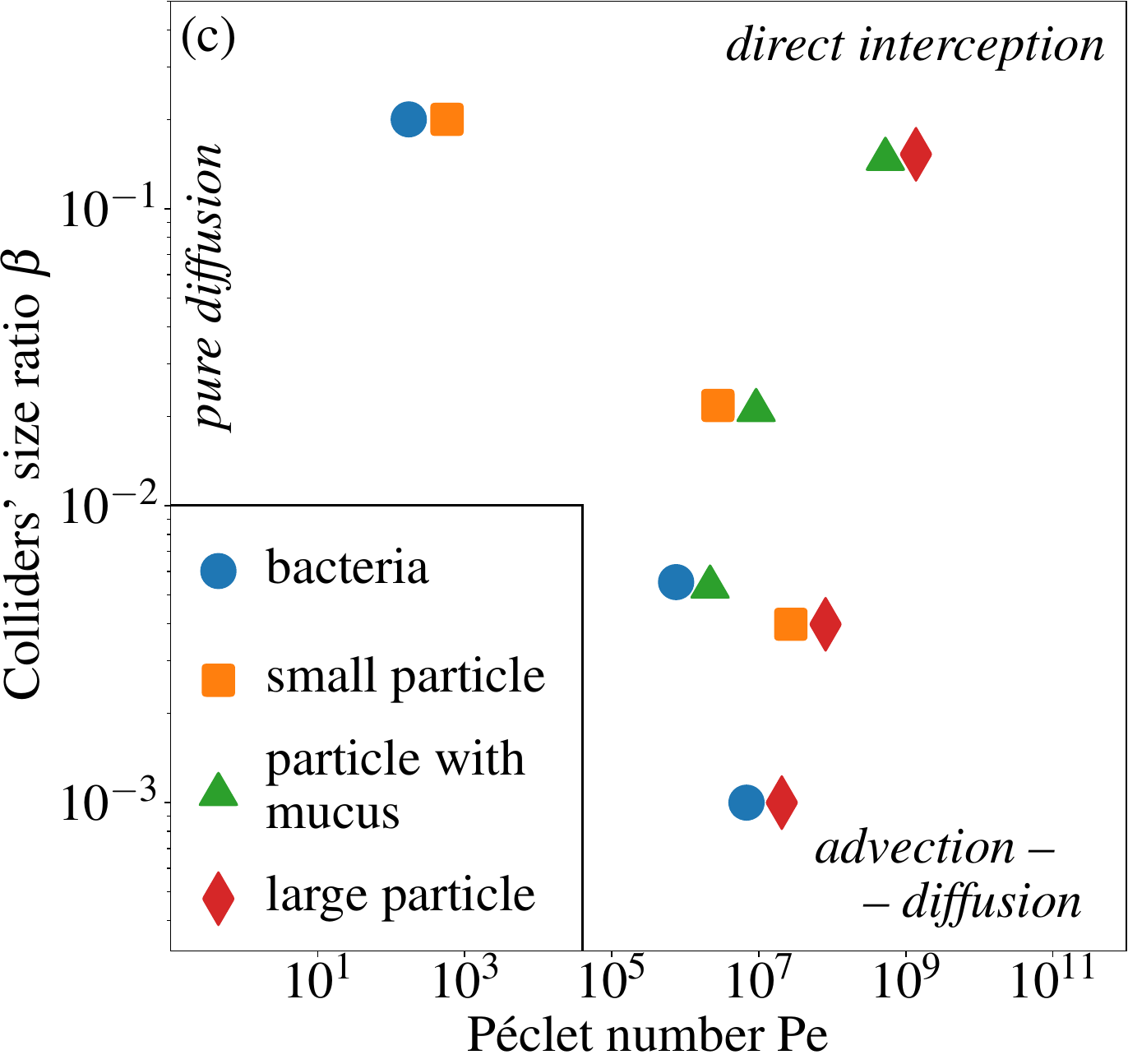}         
    \end{tabular}    
    \caption{ {\bf{The landscape of collision types in marine snow indicates the different physical encounter mechanisms at play.}}
    {(a)} Examples of different sizes and shapes of marine aggregates, imaged in situ off the coast of East Greenland. Image courtesy of E. Trudnowska, Polish Academy of Sciences. 
    {(b)} Sample image of a marine snow particle collected at 80 m below sea level, with the flow field visualised by plastic microbeads. Image by
    R. Chajwa {\it et al.}, CC BY 4.0 \citep{Chajwa_2023_arxiv}.
    {(c)} Archetypal collision types between different objects (symbols) in the parameter space of the P\'eclet number $\peclet$ and relative size $\rparticle$, based on experimental data in Table~\ref{tab:scales}. Possible collision types cover the whole space, ranging qualitatively from purely diffusive encounters, through advective-diffusive encounters, to direct (ballistic) interception. Existing collision models account for the limiting cases only.
    }
    \label{fig:snowflake-and-chmurka}
\end{figure*}

The drivers of vertical mass transport in the deep ocean, below the euphotic zone, include~\citep{Omand_2020, DeVries_2014}: sedimentation of a heterogeneous ensemble of particles \citep{Chajwa_2024,Kajihara_1971,Chase_1979,Iversen_2010}, particle remineralisation responsible for mass loss \citep{Iversen_2013,Anderson_2023,Nguyen_2022,Lambert_2019,Kiorboe_2002,Alcolombri_2021}, mass gain due to aggregation \citep{Burd_2009,Jackson_1990,Gehlen_2006,Kriest_2000}, and fragmentation by multiple mechanisms~\citep{Briggs_2020,Dilling_2000}.
Most of these processes are influenced, at least in part, by particle collisions. For example, the number of bacteria that colonize a particle can be affected by encounters with free-living populations \citep{Nguyen_2022,Kiorboe_2002}, increasing its degradation. Collisions with neutrally buoyant gels can decrease the density of a particle~\citep{Alcolombri_2025}, reducing its sinking speed. On the other hand, encounters with smaller marine snow particles can increase the sedimentation speed through mass accretion \citep{Burd_2009}. These examples highlight an important class of collisions that significantly influence the fate of marine snow particles: encounters between a large particle and small suspended objects.

As a paradigm model of encounters, researchers typically consider marine snow as a spherical particle that undergoes Stokesian sedimentation and intercepts suspended objects~\citep{Friedlander_1957,Jackson_1990,Kiorboe_1998,Kiorboe_2001c,Humphries_2009,Burd_2009}. The encounter rate in such systems has been calculated using two distinct approaches.
The first approach focuses on a direct or ballistic interception with a finite interaction range~\citep{Humphries_2009,Kiorboe_1998}. This model accounts for the non-negligible size ratio of the encountered objects and is primarily used for particles with high sinking speeds because it neglects the effects of diffusion of the objects.  
The second model is based on the advection-diffusion equation, and while it accounts for diffusion and flow around a sphere, it assumes a zero interaction range (i.e., a negligible effective size of the suspended objects)~\citep{Clift_2013,Friedlander_1957,Kiorboe_2001c,Karp_1996}. This model works well for determining the concentration and flux of oxygen onto a large particle colonised by bacteria, as established both experimentally and theoretically~\citep{Kiorboe_2001b}. However, the validity of either approach is unclear in intermediate scenarios, when the size of the objects becomes significant and their diffusion cannot be neglected.
Furthermore, the two models yield asymptotically divergent predictions depending on how the size and speed of the sinking particle and the size of the intercepted objects are varied.
Consequently, it is challenging to quantify encounter scenarios in marine snow particles accurately because the particles span many orders of magnitude in size and sinking speed. Moreover, they can collide with objects that can be both diffusive and have a finite interaction length (e.g, bacteria, gels, or other smaller marine snow particles). 

Here, we quantify encounters between sinking marine snow and suspended objects as a function of four key parameters: the size and sinking speed of the marine snow and the size and diffusivity of the objects. We determine the correct asymptotics of the encounter rate of fast-sedimenting particles,  reconciling the advection-diffusion model with the direct interception model. Our model provides a practical, closed-form formula for the encounter rate between a large sinking particle and suspended objects. Our results suggest that the number of collisions between picoplankton and larger particles may have been underestimated by the direct interception model even by two orders of magnitude.

The structure of the article is as follows. First, in Sec. \ref{sec:swept-away}, we provide an outline of physical processes involved in marine snow encounters, along with representative examples of collision scenarios which warrant theoretical quantification. In Sec. \ref{sec:governing-eqs}, we present the key equations that describe encounters and define the theoretical framework. We discuss the asymptotic solutions in Sec. \ref{sec:asymptotics}, before performing the analysis of numerical solutions in Sec. \ref{subsec:simulation}. Based on the results, in Sec. \ref{subsec:flux_calculation} we present a closed-form approximation for the encounter kernel valid for intermediate collision scenarios. We discuss the relevance of the kernel for marine snow in Sec. \ref{sec:plankton}, where we set our results in the context of pico- and nanoplankton encounters. In the following Sec. \ref{sec:limitations}, we discuss the limitations and opportunities that our approach provides. We conclude the paper in Sec. \ref{sec:conclusions}.

\section{Qualitative collision mechanisms}\label{sec:swept-away}
The nature of encounters between marine snow and suspended objects ranges from purely diffusive to purely ballistic, because marine snow particles are highly heterogeneous~\citep{Trudnowska_2021}, and cover a wide range of sizes and sinking speeds. The size of marine snow particles varies in the range of $1\;\mu$m to several mm \citep{Bochdansky_2016, mcDonnell_2010, Williams_2022}, as visible in Figure~\ref{fig:snowflake-and-chmurka}(a), with sedimentation speeds from zero to several hundreds of meters per day \citep{Williams_2022}. For most particles, the Reynolds number $\reynolds$ is less than unity~\citep{Kiorboe_2001c,Alldredge_1998}. Thus, the Stokes approximation provides a suitable starting point, as confirmed by recent measurements~\cite{Chajwa_2024} of flow fields around sinking marine snow, shown in Figure~\ref{fig:snowflake-and-chmurka}(b). 

We qualitatively describe a collision of two particles moving with relative velocity $U$ and diffusion constant $D$ considering two dimensionless numbers. With $a$ denoting the effective radius of the larger particle and $b$ the interaction range between the smaller and the larger particle (such that whenever the centres of the particles are at most $a+b$ apart they collide), we define the P\'{e}clet number $\peclet$ and $\rparticle$ describing the size ratio of the colliders as 
\begin{equation}
    \peclet = \frac{U (a+b)}{D}, \qquad \rparticle = \frac{b}{a+b}.
\end{equation}
To estimate the possible values of the two parameters, we consider four illustrative actors: a small marine snow particle (smallest measured particles~\citep{mcCave_1984}); a medium-sized, mucus laden particle \citep{Chajwa_2024}; a large particle (around the 10th particle mass quantile~\citep{Iversen_2010}); and a non-motile bacterium~\citep{Kiorboe_2002}. This set covers a broad range of possible collision parameters, outlined in Table~\ref{tab:scales}. An estimate of $\peclet$ and $\rparticle$ for collisions between them is shown in Figure~\ref{fig:snowflake-and-chmurka}(c), confirming that both $\rparticle$ and $\peclet$ span several orders of magnitude.

For very small values of $\peclet$, diffusion dominates the encounter rate. In this scenario, collisions can be viewed as stochastic, where the large, slowly sedimenting particle ``bumps into'' smaller particles. We refer to this collision mode as \emph{purely diffusive}.
When the colliders' size ratio, $\beta$, is vanishingly small but $\peclet$ becomes significant (e.g., $10^6$), advection can transport new particles into the depleted region. In this way, the sedimenting particle ``bumps into'' more particles, although the encounter mechanism remains diffusive as there is no slip at the particle surface. We refer to this collision mode as \emph{advection-diffusion}.
On the right-hand side of Figure~\ref{fig:snowflake-and-chmurka}(c), for very large values of $\peclet$ (e.g., $10^9$) and a non-negligible size ratio, $\beta$ (e.g., $10^{-2}$), previous studies have assumed that the diffusion of the smaller particle becomes negligible. In such cases, collisions can be conceptualized as a large, rapidly sedimenting particle "sweeping away" stationary smaller particles. We refer to this collision mode as \emph{direct interception}.

A natural question arises concerning the range of applicability of these models. This question escapes a simplistic answer, because only the limiting cases have been analysed so far. Without a quantitative model of the encounter rate, valid across a broad range of $\peclet$ and $\rparticle$, assuming one limiting case over another may severely underestimate the encounter rates. We develop tools to resolve this tension in the remainder of this article.

\begin{table}[tbp]
\begin{tabular}{lll}
    \toprule
    Particle or object type & Radius         & Sinking rate\\
         &  [$\mu$m] &  [m/day]     \\
    \midrule  
    Large particle \citep{Iversen_2010b}             & $1000$ & $130$ \\
    Medium-sized, mucus laden particle \citep{Chajwa_2024} & $190$ & $78$ \\
    Small particle \citep{Chase_1979}                & $4$ & $1$ \\
    Non-motile bacterium \citep{Kiorboe_2002}     & $1$ & $0$ \\
    \bottomrule
\end{tabular}
\caption{Selected representative examples of particulate matter involved in marine snow encounters, with their typical size and sinking speed as reported in experimental observations.}
\label{tab:scales}
\end{table}

\section{Governing equations}\label{sec:governing-eqs}

\begin{figure*}[t]
    \centering
    \begin{tabular}{c}
         \includegraphics[width=\linewidth,valign=t]{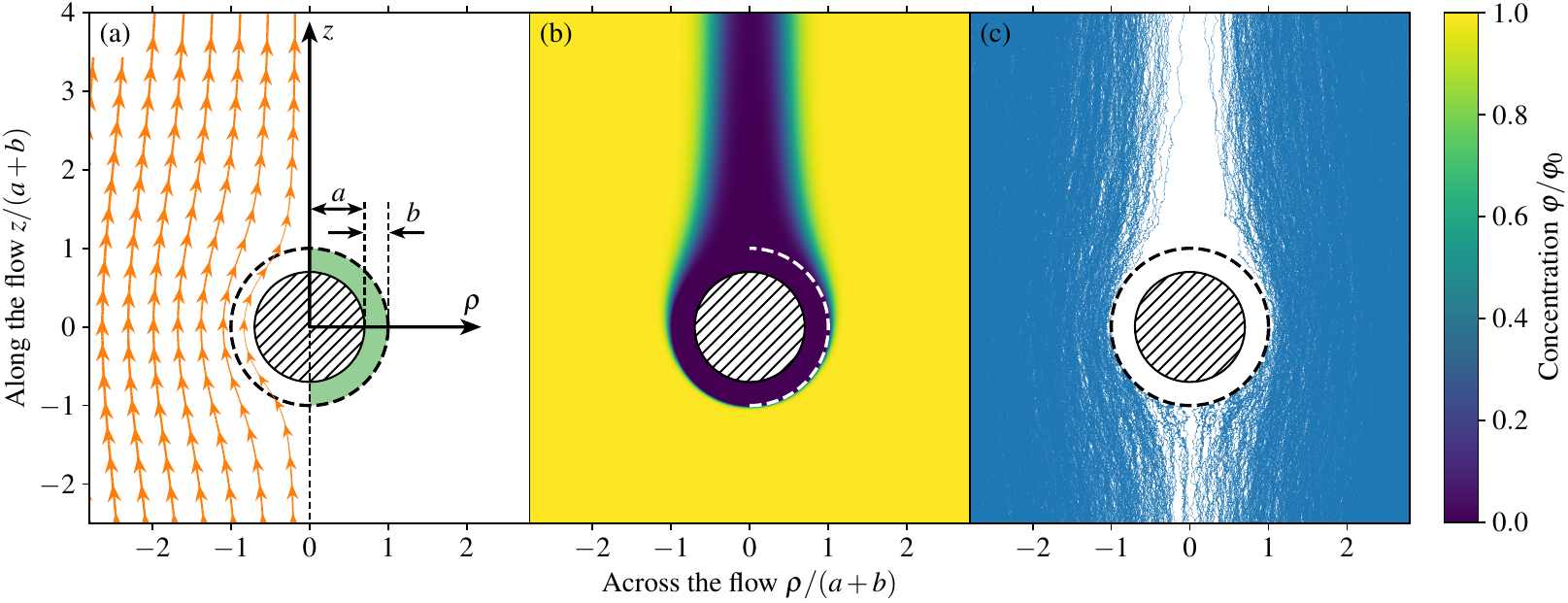}         
    \end{tabular}    
    \caption{{\bf{Sedimenting Stokesian sphere colliding with Brownian objects with non-zero interaction range}}.
    {(a)} Geometry of the collisions. Stokes flow streamlines around a particle of radius $a$ representing sedimenting marine snow, with an interaction range $b$ marked in green. The interaction radius accounts for the finite size of the suspended objects. Given the axial symmetry, we introduce sideways ($\rho$) and downstream ($z$) coordinates to parametrise the system.
    {(b)} Numerical solution of the advection-diffusion equation obtained using FEM for the steady concentration field around a sedimenting sphere (dashed area) with a size ratio $\rparticle = 0.2$ (dashed line) at $\peclet = 500$.
    {(c)} Stochastic trajectories of objects obeying the advection-diffusion equation at $\peclet = 500$. Here, $N = 3\times10^{3}$ trajectories of objects that were initially distributed uniformly on a large disk upstream from the sphere (thus more numerous further from the axis of symmetry). Trajectories that collided with the enlarged sphere (dashed) are terminated, leading to the formation of a characteristic wake free of objects behind the sphere.  
    }
    \label{fig:simulation-methods}
\end{figure*}

Consider a sphere of radius $a$ sedimenting with a velocity $U$ in a quiescent fluid. In the Stokesian regime and for an incompressible fluid, the velocity field in cylindrical coordinates $(\rho,\theta,z)$, with $\bm{U}$ aligned with the $z$ axis, is given by $\bm{u} = (u_\rho,u_z)$. The velocity components and its stream function $\psi$ expressed in the sphere frame of reference, are given by \citet{Landau_1987}
\begin{align}  
    u_\rho &= \frac{3a \rho z U}{4 R} \left( 
        \left( \frac{a}{R^2} \right)^2 - \frac{1}{R^2}  
    \right), \label{eq:eq1} \\  
    u_z &= U + \frac{3aU}{4 R} \left(
        \frac{2 a^2 + 3 \rho^2}{3 R^2}
        -\left(\frac{\rho a}{R^2}\right)^2 
        - 2
    \right),  \label{eq:eq2} \\  
    \psi &= \frac{1}{2}\, U\, \rho^2\, \left(
    1
    - \frac{3}{2} \frac{a}{R}
    + \frac{1}{2} \left( \frac{a}{R} \right)^3\;
  \right),  \\
    R^2 &= \rho^2 + z^2.
    \label{eqn:stokes_flow}
\end{align}
The streamlines of the flow field around the particle are visualised in Figure~\ref{fig:simulation-methods}(a). 
Suppose now that the suspending fluid contains small objects of effective radius $b$ which undergo diffusion relative to the large particle, with a diffusion coefficient $D$.
Whenever the distance between the centre of a small object and the centre of the sphere is smaller than $a+b$, the objects stick to the sphere and are removed from the surrounding liquid. This process can model direct contact of small spherical objects with a larger particle, but different capture mechanics can also be modelled in this fashion; one possibility is an electrostatic attraction between the particle and a small object with an effective interaction range $b$ (which can be derived by comparing the interaction potential with the typical energy of thermal fluctuations, $k_B T$). Alternatively, in the case of slip or mixed boundary conditions on the sedimenting sphere, an effective Stokes radius $a$ can be introduced and the difference between the true size and the Stokes size would give the effective interaction size. Regardless of the specific origin of such an interaction range, we focus here on the consequences of a finite interaction range described by $b$. 
We further assume that the captured particles are much smaller than the sedimenting sphere, so that their influence on the flow field around the large sphere can be neglected. In this case, the steady state concentration profile of small particles $\varphi$ is governed by the advection-diffusion equation \citep{Kiorboe_2001b} 
\begin{equation}
    0 = D \nabla^2 \varphi - \bm{u}\cdot\bm{\nabla}\varphi
    \label{eqn:advection-diffusion-raw}
\end{equation}
with the boundary condition of constant concentration, $\varphi = \varphi_0$, upstream from the ball, and $\varphi = 0$ on a sphere with an effective radius of $a+b$. 
To make Eq.~\eqref{eqn:advection-diffusion-raw} dimensionless, we choose the time scale to be $(a+b) / U$, use $a+b$ as the length scale, and $\varphi_0$ as the concentration scale and arrive at the dimensionless form
\begin{equation}
    0 =  \nabla^2 \varphi - \peclet\,(\bm{u}\cdot\bm{\nabla}\varphi),
    \label{eqn:advection-diffusion}
\end{equation}
with the boundary conditions for concentration
\begin{equation}
    \begin{split}
        &\varphi(r = 1) = 0,\\
        &\varphi(r \to \infty) = 1.
    \end{split}
    \label{eqn:boundary_condition}
\end{equation}
The dimensionless velocity field, Eqs.~\eqref{eq:eq1}-\eqref{eq:eq2}, can be written in terms of the rescaled large particle radius $\alpha = a/(a+b)$ as
\begin{equation}
    \begin{split}
        u_\rho &= \frac{3 \alpha \rho z}{4 R} \left( 
        \left( \frac{\alpha}{R^2} \right)^2 - \frac{1}{R^2} 
    \right), \\  
    u_z &= 1 + \frac{3\alpha}{4 R} \left(
        \frac{2 \alpha^2 + 3 \rho^2}{3 R^2}
        -\left(\frac{\rho \alpha}{R^2}\right)^2 
        - 2
    \right). \\  
    \end{split}
\end{equation}
Crucially, the boundary conditions of no slip and no concentration are imposed here on different surfaces, with $\bm{u}(r=\alpha)=0$, rather than the classical case, where $\bm{u}(r=1)=0$.

We now calculate the mass of small particles intercepted by the large particle per unit time. Expressing~\eqref{eqn:advection-diffusion} as $0 = \bm{\nabla} \cdot \bm{J}$, we can write the particle current $\bm{J}$ as
\begin{equation}
    \bm{J} = \peclet \, \bm{u} \varphi - \bm{\nabla} \varphi,
\end{equation}
where we used the fact that the flow is incompressible, $\bm{\nabla}\cdot\bm{u}=0$. 
The total, dimensionless flux of particles onto the sphere, $\Phi^*$, is given by
\begin{equation}
    \Phi^* = \oint_{\textrm{sphere}} \bm{J} \cdot \dd \bm{S}.    
    \label{eqn:flux_def}
\end{equation}
Using the Stokes theorem, and given the lack of source terms in Eq.~\eqref{eqn:advection-diffusion}, the integration surface in Eq.~\eqref{eqn:flux_def} can be changed to any other surface enclosing the sphere. When the integration surface is taken as a large cylinder that extends far from the sphere, we have $\partial_z \varphi \ll \bm{u} \varphi$, and we can avoid calculating the gradient in the numerical evaluation of $\bm{J}$. The dimensional flux is calculated as $\Phi = \varphi_0 D (a+b)\Phi^*$. With precise definitions, we are ready to discuss different approximations used to compute $\Phi$ for given values of $\peclet$ and $\beta$.

\section{Divergent asymptotics of the limiting cases}\label{sec:asymptotics}

Before solving Eq.~\eqref{eqn:advection-diffusion} in the most general case ($\peclet\geq 0$, $\rparticle\geq 0$), we first briefly discuss its important limiting cases, namely the widely used direct interception limit and the advection-diffusion description with zero interaction range. We also highlight important scenarios under which these limiting cases break down and yield asymptotically divergent predictions.

In the absence of flow ($\peclet \to 0$,  \emph{purely diffusive encounters}), Eq.~\eqref{eqn:advection-diffusion} becomes a spherically symmetric Laplace's equation, giving the classical expression for the diffusive flux $\Phi_\textrm{D}$ onto a sphere \citep{Karp_1996}
\begin{equation}
    \Phi_\textrm{D} = 4 \pi D (a+b)\varphi_0.
    \label{eqn:diffusive_flux}
\end{equation}

When flow is present ($\peclet>0$, \emph{advective-diffusive encounters}) it is customary to normalize $\Phi$ by this limiting case to obtain the dimensionless quantity proportional to the encounter rate. This ratio defines the  Sherwood number~\citep{Karp_1996}, given by
\begin{equation}
    \sherwood =\frac{\Phi}{\diffusiveflux}.
    \label{eqn:sherwood_definition}
\end{equation}
Note that there is no consensus on the definition of $\sherwood$ and $\peclet$ numbers in the literature and the definitions might differ by a factor of two. 

For an intermediate range of $\peclet$ values and in the limit of a zero interaction range ($\rparticle\to 0$), Eq.~\eqref{eqn:advection-diffusion} also describes heat transfer to or from a sphere in a slowly flowing fluid. In the limit of small or higher Peclet, analytical solutions of this problem were obtained using perturbative methods \citep{Brunn_1982,Gupalo_1972,Rimmer_1968,Acrivos_1962,Acrivos_1965,Bell_2013}. Intermediate cases require numerical simulations, pursued since the early work of \citet{Friedlander_1957}. To date, numerical solutions of heat transfer between a sphere immersed in a colder, flowing liquid are well established \citep{Feng_2000,Clift_2013,Westerberg_1990}, and a simple, closed form approximation of numerical results provided by \citet{Clift_2013} gives the Sherwood number $\sherwood_{\textrm{Cl}}$ in terms of $\peclet$ with great accuracy
\begin{equation}
   \sherwood_{\textrm{Cl}} =  \frac{\cliftflux}{\diffusiveflux}= \frac{1}{2}\left(1+(1+2\peclet)^{1/3}\right).
    \label{eqn:Sh_clift}
\end{equation}
These are in agreement with experimental results considering a heated sphere in slowly flowing fluid~\citep{Kramers_1946} and absorption of ions by a conducting sphere~\citep{Kutateladze_1982}. However, we stress that all of these approaches assume a zero interaction range $\rparticle\to0$.

When advection dominates over diffusion and the interaction range is finite ($\peclet \to \infty$, $\beta> 0$, \textit{direct interception}), a different family of approximations has been derived. In this regime, the Laplacian term in Eq.~\eqref{eqn:advection-diffusion} can be neglected, and the stationary solution takes only two values: $0$ inside a critical streamline and $1$ outside of it. To determine $\Phi$, one simply calculates the cross section of the critical stream tube in the far-field flow in Eq.~\eqref{eqn:stokes_flow}. This gives expression for \emph{direct interception} flux $\Phi_{\textrm{A}}$ as~\citep{Friedlander_1957}
\begin{equation}
    \Phi_{\textrm{A}} = U \pi b^2 \frac{3-\rparticle}{2}\varphi_0.
    \label{eqn:Phi_advective}
\end{equation}

However, predicting encounters for fast sinking particles based on current models depends strongly on how the limit of high $\peclet$ is approached, with the direct interception model yielding asymptotically divergent predictions from the advection-diffusion model. Intuitively, consider the volume $V_\textnormal{c}$ swept clean by a sedimenting particle over a fixed travel distance $\Delta Z$. For a given diffusion coefficient $D$, Eq.~\eqref{eqn:sherwood_definition} predicts that the volume is
\begin{equation}
    V_\textnormal{c}=4\pi a^2  \, \frac{\Delta Z}{\peclet}\sherwood,
    \label{eqn:volume_swept}
\end{equation}
which is vanishingly small for large $\peclet$ because, from Eq.~\eqref{eqn:Sh_clift}, $\sherwood_{\textrm{Cl}} \approx \left( \peclet/4 \right)^{1/3}$ as $\peclet\to\infty$.
In other words, the advection-diffusion model in Eq.~\eqref{eqn:Sh_clift} predicts that marine snow stops intercepting new particles when it becomes large enough.
In contrast, the cleared volume predicted by the direct interception model (Eq.~\eqref{eqn:Phi_advective}) is always finite and given by $V_\textnormal{c}=\pi b^2 (3-\rparticle)  \Delta Z / 2$. This divergence of the predicted encounter rate between the two models arises because the advection-diffusion model assumes an infinitely small interaction range ($\rparticle = 0$).

Thus, for any interaction range, however small, the interaction range cannot be neglected at very large values of $\peclet$. The true asymptotics of $\sherwood(\peclet)$ are determined by $\rparticle$ and, eventually, to maintain a constant, non-zero $V_\textnormal{c}$ from Eq.~\eqref{eqn:volume_swept}, $\sherwood(\peclet)$ should scale as $\sherwood(\peclet) \sim \peclet$.
To explore this discrepancy, we next calculate $\sherwood$ numerically in the general case~($\peclet\geq 0$ and $\rparticle\geq 0$).

\begin{figure}[h]
    \centering
    \begin{tabular}{c}
         \includegraphics[width=\linewidth,valign=t]{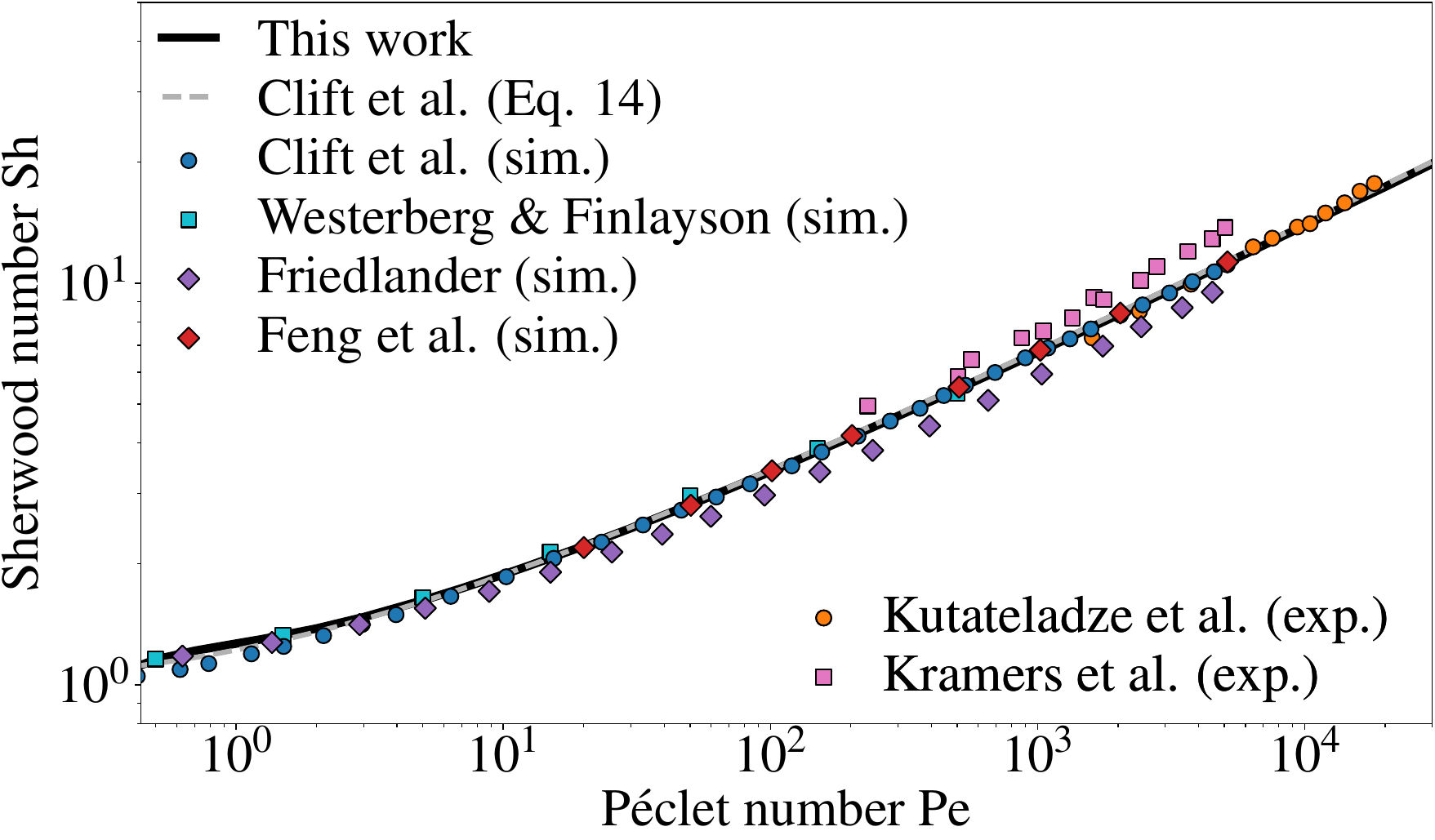}         
    \end{tabular}    
    \caption{{\bf{Validation of our numerical results in the case of zero interaction range}.}
    Comparison of our simulations with earlier experimental \citep{Kramers_1946, Kutateladze_1982} and numerical \citep{Friedlander_1957,Westerberg_1990,Feng_2000,Clift_2013} solutions of the advection-diffusion problem around a sphere in Stokes flow (thus with a zero interaction range, $\rparticle=0$). 
    Closed-form approximation of \citet{Clift_2013} is shown as dashed line.
    Results from our numerical model show excellent agreement with earlier numerical works (excluding the work of \citet{Friedlander_1957} which is an outlier). Deviations from experimental data are likely driven by finite Reynolds number effects (\citet{Kramers_1946} measured $\sherwood$ with $0.42<\reynolds<26.5$). Different definitions of P\'{e}clet and Sherwood numbers were harmonised.     
    }
    \label{fig:literature-comparison}
\end{figure}

\section{Numerical investigation}\label{sec:numerical}

\subsection{Simulation methods}\label{subsec:simulation}

Equation~\eqref{eqn:advection-diffusion} can be solved directly using finite element methods (FEM) for moderate values of P\'{e}clet number such as $1 < \peclet < 10^{6}$. We used the Python package \code{scikit-fem} \citep{skfem_2020}, working in cylindrical symmetry. We expressed Eq.~\eqref{eqn:advection-diffusion} in a weak form and obtained solutions for the concentration profiles around the sedimenting sphere, as shown in Figure~\ref{fig:simulation-methods}(b). We provide more details on this method of solution in Figure~S1 in Supplementary Materials (SM) and present convergence tests in Figure~S2 and Figure~S3 in SM. The concentration profiles were post-processed to obtain $\sherwood$. We did it in two ways to further validate the numerical solution. In the first approach, we calculated $\Phi$ by integrating $\peclet\, \varphi(\rho) \bm{u}(\rho)$ far downstream of the sphere. In the second approach, we calculated $\Phi$ by integrating $(\peclet\, \varphi \bm{u} - \nabla \varphi) \cdot \dd \bm{S}$ on the surface of the capturing sphere. Whenever the simulation box was sufficiently big, these methods were in agreement. 
For smaller values of $\peclet$ (e.g. $\peclet \sim 0.1$) FEM becomes impractical because it requires a very large simulation box to set the boundary conditions properly. However, this regime is well described by $\sherwood_{\textrm{Cl}}$, defined in Eq.~\eqref{eqn:Sh_clift}.

We validated the FEM approach with solutions and experiments available in the literature for $\rparticle=0$. The comparison of our numerical results for $\sherwood(\peclet)$ is shown in Figure~\ref{fig:literature-comparison}. After harmonising the definitions of $\sherwood$ and $\peclet$ between different sources and this publication, we see that our simulations are in good agreement with previous numerical and experimental works, and agree with the approximate relationship of Eq.~\eqref{eqn:Sh_clift}. 

\begin{figure}[htbp]
    \centering
    \begin{tabular}{c}
         \includegraphics[width=\linewidth,valign=t]{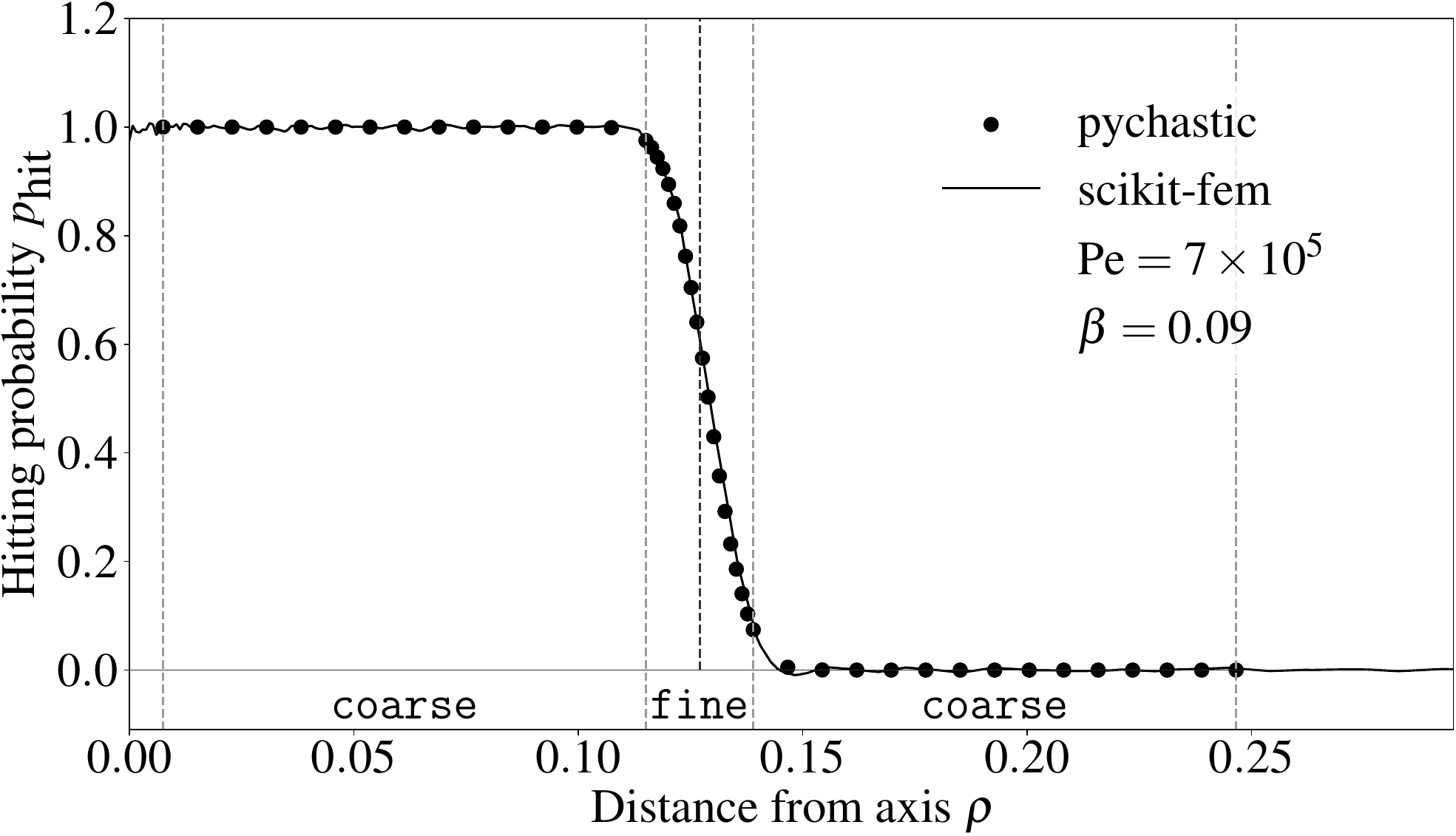}         
    \end{tabular}    
    \caption{{\bf{Hitting probability of diffusive objects with a non-zero interaction range, calculated using SDE and FEM simulations}.} {Numerical calculations for a representative, intermediate value of $\peclet$ and $\beta$ show good agreement between the FEM (\code{scikit-fem}) and SDE (\code{pychastic}) methods. In the SDE approach, each point was computed using $N=10^4$ trajectories. To reduce computational time, a set of 50 initial conditions was chosen. In a region near the critical streamline, sampling density was increased (more details in Supplementary Materials, in Figure~S4 and Figure~S5). The profile of hitting probability was used to calculate the total flux onto the particle using Eq.~(\ref{eq:phi_sde}).}}
    \label{fig:fem_vs_pychastic}
\end{figure}

For even higher $\peclet$ numbers, FEM calculations proved ineffective due to a numerical stability problem: for very large values of $\peclet$, ringing artifacts appear near sharp gradients which follow the streamlines. Since the calculations for $\peclet > 10^7$ are key to understanding marine snow collisions, as seen in Figure~\ref{fig:snowflake-and-chmurka}(c), we employed another simulation method.

Given that Eq.~\eqref{eqn:advection-diffusion-raw} is of Fokker-Planck type, it is possible to obtain integrals of $\varphi$ by simulating the corresponding \Ito{} stochastic differential equation (SDE) with appropriate initial and boundary conditions~\cite{van_Kampen_1992}. The simulated trajectories are used to calculate the hitting probability for a given initial position, which quantifies the number of small objects absorbed by the large particle. The total flux is then obtained by integrating the hitting probability over a disk placed upstream of the particle.

Specifically, the positions of suspended objects, $\bm{q}$, are described by the equation
\begin{equation}
    \dd \bm{q} = \bm{u} \dd t + \sqrt{2 D}\, \dd \bm{W}_t,
\end{equation}
with $\bm{q}(t=0) = \bm{q_0}$, and where $\bm{W}_t$ is the Wiener process.  Scaling the lengths by $(a+b)$ and the time by $(a+b)/U$ leads to the dimensionless expression
\begin{equation}
    \dd \bm{q} = \bm{u} \dd t + \sqrt{\frac{2}{\peclet}}\, \dd \bm{W}_t.
\end{equation}
We simulated $N = 10^4$ particles for each of 50 starting points, $\bm{q_0} = (x_0,0,-h)$, with $h = 5$, using the Python package \code{pychastic}~\citep{Waszkiewicz_2023,pychastic_codebase}. The typical resulting trajectories are shown in Figure~\ref{fig:simulation-methods}(c). The trajectories were then used to calculate the hitting probability $p_{\mathrm{hit}}(x_0)$ as a function of $x_0$ for an ensemble of initial locations $x_0$. Specifically, $p_{\mathrm{hit}}(x_0)$ is the fraction of trajectories that approach the particle at a distance smaller than 1. The flux $\Phi$ was calculated as
\begin{equation} \label{eq:phi_sde}
    \Phi =\varphi_0 U(a+b)^2 \int_0^\infty 2 \pi \rho \, p_{\mathrm{hit}}(\rho) \, u_z(\rho) \, \dd \rho.
\end{equation}
To efficiently calculate the integral, we sampled $p_{\mathrm{hit}}$ on a non-uniform grid (Figure~\ref{fig:fem_vs_pychastic}; for details, see discussion of Figure~S4 in SM) and used linear interpolation of $p_{\mathrm{hit}}$. We note that Eq.~\eqref{eq:phi_sde} includes the contribution of the advective flux ($\bm{u}\varphi$) in Eq.~\eqref{eqn:flux_def} and ignores the diffusive component ($D\nabla\varphi$); This approximation utilizes the fact that $\pd_z\varphi$ is negligible far downstream from the sphere.

In summary, FEM is an effective tool for simulating the dynamics of small objects for $\peclet<10^{7}$, while in the high-Pe regime, $\peclet>10^5$, SDE trajectories are increasingly more practical due to faster convergence. We used the intermediate regime, where the two methods overlap, as a test ground to compare their results. In Figure~\ref{fig:fem_vs_pychastic} we present an estimation of the hitting probability profile for $\peclet=7\times10^5$ and $\beta=0.09$ using FEM and SDE. In this case, both methods converge. The details of the compatibility tests performed are included in the SM, in Figure~S5 and Figure~S6. 

\subsection{Flux calculation - results}\label{subsec:flux_calculation}

\begin{figure}[tbp]
    \centering
    \includegraphics[width=\linewidth,valign=t]{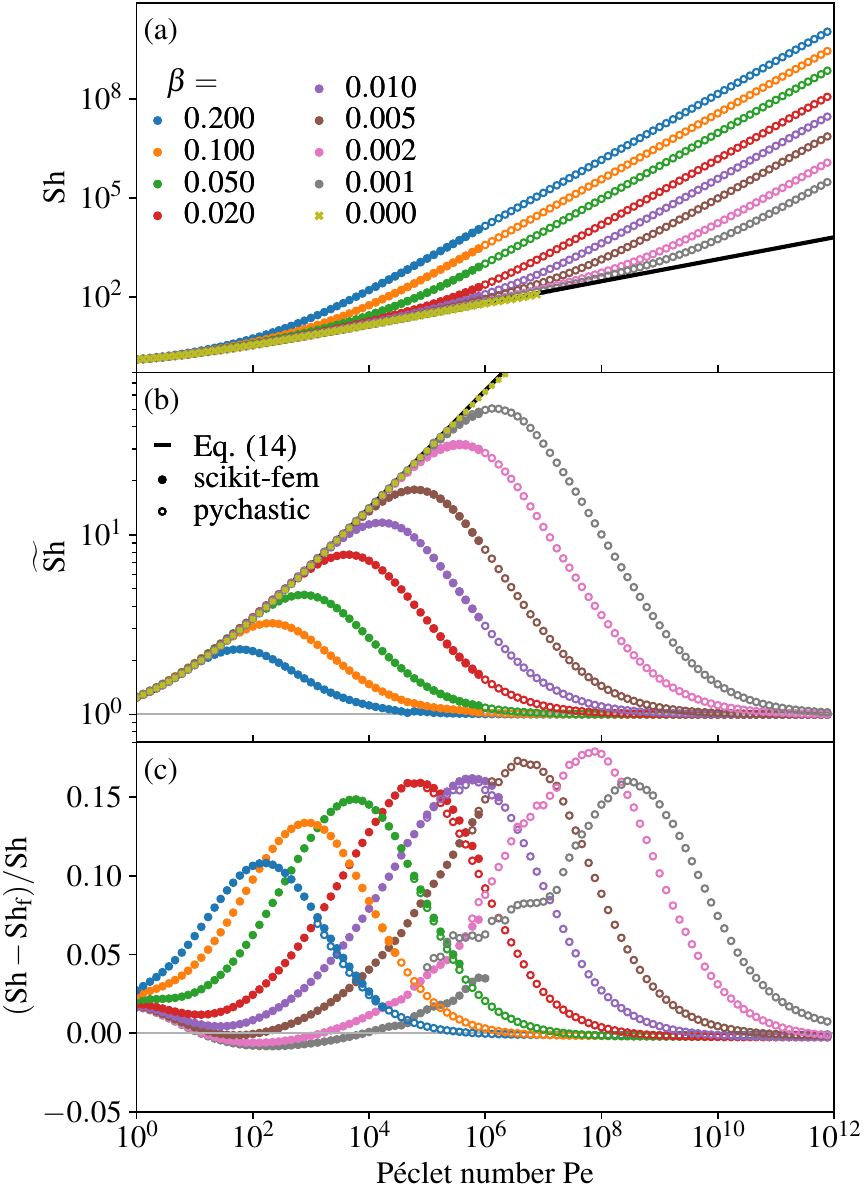} 
    \caption{{\bf Sherwood number for varying $\peclet$ and $\rparticle$, obtained from numerical simulations.} Filled dots correspond to finite element method (FEM) using the \code{scikit-fem} package, empty dots correspond to solutions of the stochastic differential equation (SDE) using the \code{pychastic} package. Color denotes the value of $\rparticle$. {(a)} Sherwood number, $\sherwood$, as a function of $\peclet$. The solutions range from ignoring the radius of objects (aligning with the solution of \citet{Clift_2013}) to ignoring diffusion of objects (parallel straight lines in the high-$\peclet$ limit).
    {(b)} Results of the same calculation presented in terms of the modified Sherwood number, $\modsherwood$, defined in Eq.~\eqref{eq:modified-sh} . With this parametrisation, all solutions approach $1$ as $\peclet \to \infty$.
    {(c)} Relative error between numerical results for $\sherwood$ and the analytical approximation $\sherwood_{\textrm{f}}$, proposed in Eq.~\eqref{eqn:approximation}.}
    \label{fig:sh_vs_pe}
\end{figure}

The calculated flux $\Phi$ reveals a different asymptotic scaling with large $\peclet$ for objects with a non-zero interaction range ($\rparticle>0$) compared to objects with zero interaction range ($\rparticle=0$).
We calculated $\Phi$ for different values of $\peclet$ and $\rparticle$ and compared our calculations with the solution from \citet{Clift_2013} ($\rparticle=0$), given by Eq.~\eqref{eqn:Sh_clift}  (Figure~\ref{fig:sh_vs_pe}(a)). We find that, for small $\peclet$, all solutions with different $\rparticle$ converge to the relation $\sherwood_{\textrm{Cl}}$. However, for higher $\peclet$, the solutions diverge significantly from the curve with $\rparticle=0$ after a transition region, eventually reaching an asymptote different from $\sherwood_{\textrm{Cl}} \approx \left( \peclet/4 \right)^{1/3}$. Instead, we observe that the solutions for $\rparticle\neq 0$ approach the asymptotics of $\sherwood \sim \peclet$ in the high-$\peclet$ limit. 

To better capture the transition between the diffusion and direct interception regimes, we introduce the modified Sherwood number $\modsherwood$ as
\begin{equation} \label{eq:modified-sh}
    \modsherwood = \frac{\Phi}{\diffusiveflux+\advectiveflux}.
\end{equation}
This extended definition ensures that all solutions for $\rparticle>0$ tend to a finite limit as $\peclet \to \infty$. As shown in Figure~\ref{fig:sh_vs_pe}(b), all numerical solutions for $\beta>0$ eventually detach from $\sherwood_{\textrm{Cl}}$ at a given $\peclet$ and tend to 1, thus transitioning from advection-diffusion with zero interaction range to pure direct interception.

\begin{figure*}[htbp]
    \centering
    \includegraphics[width=\linewidth,valign=t]{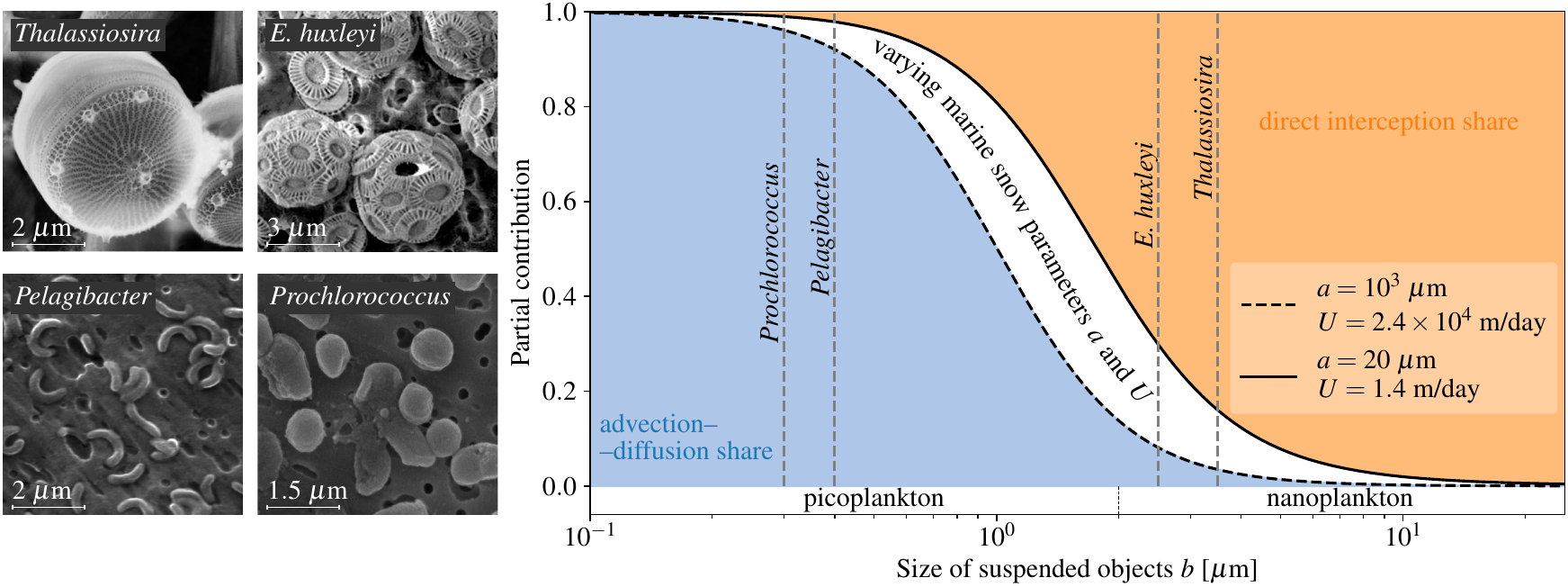} 
    \caption{
    \textbf{
    Advection-diffusion vs. direct interception for pico- and nanoplankton.}
    Partial contribution to the total flux from $\cliftflux$ (denoted as the advection-diffusion share) and $\advectiveflux$ (denoted as direct interception share). 
    The white region represents a family of curves reflecting the partial contribution generated by varying $\Delta \rho$ and a ($30<\Delta \rho <200$ and $20\;\mu\textrm{m}<a<10^3\;\mu\textrm{m}$) in the Stokes law. An increase of $\Delta \rho$ and $a$ leads to an increase of $\advectiveflux$ in comparison to $\cliftflux$. See Figure S8 in SM for parameterization using smaller density differences. The vertical dashed lines represent two examples of non-motile pico- and nanoplankton each: \emph{Prochlorococcus} (Image by:
    Anne Thompson, public domain~\citep{Thompson_2009}) and \emph{Pelagibacter} (Image by:
    Laura Steindler {\it et al.}, CC BY 4.0~\citep{Steindler_2011}) as well as \emph{E. huxleyi} (Image by: Maria Iglesias-Rodriguez {\it et al.}, CC BY 4.0~\citep{Iglesias_2017}) and \emph{Thalassiosira} (Image reproduced from \citep{Sumper_2008} with permission).}
    \label{fig:bacterial_encounters}
\end{figure*}

We provide two ways of incorporating the improved estimates of $\Phi$ for future research. First, we developed a Python package \code{pypesh} accompanying this manuscript which, for given $\peclet$ and $\rparticle$, calculates $\sherwood$ as output interpolated from the calculations from Figure~\ref{fig:sh_vs_pe}(a).
Second, we propose a closed-form approximation of the numerical results. We approximate $\Phi$ as the sum of $\cliftflux$ and $\advectiveflux$
\begin{align}
\begin{split}
    \sherwood\approx\sherwood_{\textrm{f}} &=(\cliftflux+\Phi_{\textrm{A}})/\diffusiveflux= \\
    & =\sherwood_{\textrm{Cl}} + \frac{U \pi b^2}{4 \pi D (a+b) } \frac{3-\rparticle}{2} = \\
    & = \frac{1}{2}\left(1+(1+2\peclet)^{1/3}+ \peclet\:\frac{\rparticle^2(3-\rparticle)}{4}\right).
\end{split}
\label{eqn:approximation}
\end{align}
Figure~\ref{fig:sh_vs_pe}{(c)} shows the relative difference between the numerically calculated value of $\sherwood$ and the approximation of Eq.~\eqref{eqn:approximation}. We see that $\sherwood_\text{f}$ underestimates the numerical value by at most 20\% , with the maximum of deviations shifting towards higher $\peclet$ for smaller values of $\rparticle$. In the solution for $\rparticle=0.001$, a delicate discrepancy occurs, between the two numerical schemes, caused by numerical errors of the estimated $\sherwood$. This error, however, is smaller than 5\%. In summary, the formula in Eq.~\eqref{eqn:approximation} provides a closed-form approximation of the flux as a function of $\peclet$ and $\rparticle$.
 
\section{Ecological implications for pico- and nanoplankton}\label{sec:plankton}

We now apply our results to environmentally realistic scenarios. We now quantify the relative contribution of advection-diffusion (Eq.~\eqref{eqn:Sh_clift}) and direct interception (Eq.~\eqref{eqn:Phi_advective}) to the total encounter kernel (Eq.~\eqref{eqn:approximation}) as a function of the size of the suspended objects in Figure~\ref{fig:bacterial_encounters}. 
We assume the smaller objects to be neutrally buoyant and have a diffusion coefficient described by the Stokes-Einstein relationship
\begin{equation}
    D=\frac{k_{\textrm{B}}T}{6\pi\mu b},
    \label{eqn:diff_const}
\end{equation}
where $k_{\textrm{B}}=1.38\times10^{-23}\;\textrm{J}/\textrm{K}$ is the Boltzmann constant, $T=277\;\textrm{K}$ is the approximate temperature of seawater and $\mu = 1.6\times10^{-3}\;\textrm{Pa}\,\textrm{s}$ is the dynamic viscosity of seawater. To calculate $D$ for small objects, we make a common assumption that the interaction range and the size of the object are identical~\citep{Humphries_2009, Kiorboe_2002, Jackson_1990, Burd_2009, Alcolombri_2025}.

We assume for simplicity that larger particles sink according to the Stokes' law
\begin{equation}
    U = \frac{2 g \Delta \rho}{9\mu}a^2,
\end{equation}
where $\Delta \rho$ is the density difference between water and the particles, and $g = 9.81\;\textrm{m}/\textrm{s}^2$ is the gravitational acceleration. We consider $ a $ from $ 20 \; \mu\textrm{m} $ to $ 10^3 \; \mu\textrm{m} $ to cover a realistic range of marine snow sizes. To mimic the variability in sinking velocities, we consider two ranges of $\Delta \rho$. First, we set $ 30 \; \textrm{kg}/\textrm{m}^3 <\Delta \rho<200 \; \textrm{kg}/\textrm{m}^3$  based on previously reported values~\citep{Chase_1979,mcDonnell_2010}. Second, we consider much smaller values of $\Delta\rho$ to represent slowly sinking particles.

Our formula implies that encounters between sinking marine snow and submicrometer-sized suspended objects, such as non-motile picoplankton, are driven primarily by advection-diffusion, whereas the capture mechanism for larger objects, such as non-motile nanoplankton, depends sensitively on the properties of the sinking marine snow. Figure~\ref{fig:bacterial_encounters} shows the relative contribution of the two terms (advective-diffusive in blue and direct interception in orange) to the total flux as a function of the size of suspended objects for the case of marine snow density differences in the range of $ 30 \; \textrm{kg}/\textrm{m}^3 <\Delta \rho<200 \; \textrm{kg}/\textrm{m}^3$. In this regime, the share of advection-diffusion vs. direct interception contributions takes a sigmoidal shape, centred at ca. $1.5\; \mu \textrm{m}$, a location that is close to the traditional size threshold separating marine microorganisms into picoplankton and nanoplankton~\cite{Omori_1992}. For a given size of suspended objects, the contributions of $\cliftflux$ and $\advectiveflux$ are weakly dependent on the size of the marine snow particle $a$ and vary mainly with $\Delta \rho$ (with the range of relative shares of $\cliftflux$ and $\advectiveflux$ constrained by the white region; see Figure~S7 in the SM for more details). For example, when the smaller object has a radius of $b \approx 1 \; \mu\textrm{m}$, the share of advection-diffusion is between $50\%$ and $83\%$, and the share of direct interception is between $17\%$ and $50\%$, respectively, depending on the density of the particle. On the other hand, for $b \approx 2.5 \; \mu\textrm{m}$, direct interception contributes between $64\%$ and $91\%$ of the total flux. For excess density of marine snow less than $ \Delta \rho = 30\;\textrm{kg}/\textrm{m}^3 $, the white region in Figure~\ref{fig:bacterial_encounters} shifts towards larger values of $b$, while maintaining its sigmoidal shape. In the case of $ \Delta \rho \approx 10^{-2}\;\textrm{kg}/\textrm{m}^3 $, which corresponds to $ U = 1.2\;\textrm{m}/\textrm{day} $ for $ a = 10^3\;\mu\textrm{m} $ (such slowly sinking particles have also been observed in situ~\citep{Williams_2022}), the 50\% threshold on the right-hand side of Figure~\ref{fig:bacterial_encounters} occurs at $ b \approx 12\;\mu\textrm{m} $ (see the Figure~S8 in SM for more details). This estimation shows that as the particle slows down, direct interception may underestimate the encounter rate with suspended objects as large as nanoplankton.

\section{Discussion}\label{sec:limitations}

Estimating microhydrodynamic encounter rates is crucial for quantifying the time scales of microscale interactions~\cite{Kiorboe_2008,Burd_2009,Slomka_2023}. In the context of marine snow particles, collisions with small suspended objects shape the dynamics of carbon export to depth by mediating the colonization of the particles by microorganisms or the acquisition of additional ballast. 

Such collisions are often quantified by the zero-interaction model ($\beta = 0$)~\citep{Kapellos_2022, Nguyen_2022} or by neglecting the diffusion of objects~\citep{Humphries_2009, Krishnamurthy_2023}. However, as we have shown in Sec. \ref{sec:asymptotics}, the quantification of the encounter process is hampered by the different asymptotics of these two frequently used models. Through Eq.~(\ref{eqn:approximation}), our work provides a simple method that accurately describes both regimes and the intermediate scenarios. With an increasing number of studies determining the size distribution of marine snow particles~\citep{Jackson_1997, Bochdansky_2016, Cavan_2018, mcCave_1984} and their sinking speeds~\cite{Williams_2022,Chajwa_2024}, we expect our kernel to become a useful tool for assessing the importance of different encounter mechanisms on the overall sedimentation flux \cite{Burd_2009,Nguyen_2022}.

Importantly, we showed that considering only $\peclet$ is insufficient to rule out the contribution of diffusion to generating encounters. For example, as shown in Figure~\ref{fig:sh_vs_pe}, for $\peclet=10^6$, the direct interception model works very well when $\rparticle=0.02$ (e.g., in the case of a large particle with $a\approx175\;\mu\textrm{m}$, $U\approx20\;\textrm{m}/\textrm{day}$, capturing small diatoms, such as \textit{Thalassiosira}, with $b\approx3.5\;\mu\textrm{m}$). In contrast, advection-diffusion provides an almost exact result for $\rparticle=0.001$ (e.g., in the case of a large particle with $a\approx300\;\mu\textrm{m}$, $U\approx100\;\textrm{m}/\textrm{day}$ capturing a cyanobacteria \textit{Prochlorococcus}, $b\approx0.3\;\mu\textrm{m}$), implying that relying on direct interception alone underestimates the encounter rate by a factor of 200 in this case. Similarly, our work implies that the scavenging of neutrally buoyant biogels may occur more frequently than previously thought, possibly enhancing the slow-down of sinking particles \cite{Alcolombri_2021}, or the formation of mucus comet tails \cite{Chajwa_2024}. 

Our numerical results highlight the need to validate the {\it ad hoc} summation of encounter kernels. Several heuristic approaches have been proposed to describe the interception of small objects by a large sinking particle, including using the direct interception kernel alone (Eq.~\eqref{eqn:Phi_advective})~\cite{Kiorboe_2001a}, the sum of the direct interception and purely diffusive kernels (Eqs.~\eqref{eqn:Phi_advective} and~\eqref{eqn:diffusive_flux})~\cite{Jackson_1990,Burd_2009}, or the sum of direct interception (Eqs.~\eqref{eqn:Phi_advective} and an asymptotic variant of Eq.~\eqref{eqn:Sh_clift}, valid for sufficiently large $\peclet$~\cite{Shimeta_1991,Shimeta_1993}). Here, we have shown that a specific choice of Eq.~\eqref{eqn:approximation}, which accounts for the summation of the direct interception and advective-diffusive kernels (Eqs.~\eqref{eqn:Phi_advective} and~\eqref{eqn:Sh_clift}), provides an accurate description of encounter processes valid for all regimes. In the general case, with other collision mechanisms involved~\citep{Jackson_1990, Kriest_2000, Gehlen_2006, Burd_2009, Aumont_2015, Stock_2020}, our results call for a critical assessment of the assumption of kernel additivity. Beyond models of marine snow collisions, the kernel in Eq.~\eqref{eqn:approximation} may find applications in particle-laden flows relevant to atmospheric phenomena~\cite{Rosa_2013} and industrial processes such as flotation~\cite{Jiang_2025}.

Our model and its potential applications are subject to four primary limitations. First, larger, denser marine snow particles can achieve sedimentation speeds that invalidate the Stokes flow assumption~\citep{Kiorboe_2001c,Alldredge_1998}, and require extending the non-zero $\reynolds$ model of \citet{Humphries_2009} to include the diffusion of objects. Second, our approach neglects hydrodynamic interactions that become relevant when the sizes of colliders are comparable, and can be accounted for by including the distance-dependent hydrodynamic mobility of two spheres~\cite{Jeffrey_1984,Lisicki_2014}. Third, we have assumed non-motile objects. Motility can be incorporated using effective diffusion~\cite{Lambert_2019} in the case of large marine snow, whereas encounters with smaller marine snow must take into account the reorienting effects of the shear profile on motile cells~\cite{Slomka_2020}. Finally, we assumed effective spherical shapes of the sinking particles. While marine snow particles have irregular shapes and calculating encounter rates for a given shape of a collider remains largely an open question, we consider here an effective hydrodynamic radius and expect that the primary role of the shape is to determine the particle's sinking speed. In amorphous cases, we expect the demarcation between advective-diffusive and ballistic encounters predicted by Eq.~\eqref{eqn:approximation} to hold more broadly.

\section{Conclusions}\label{sec:conclusions}

In this study, we theoretically and numerically addressed the problem of quantifying encounters between a large sinking sphere and suspended objects, depending on the size and sinking speed of the large sphere, as well as the size and diffusivity of the objects. Using advection-diffusion simulations, based on the finite element method (FEM) and stochastic differential equation (SDE) approaches, we quantified the encounter rate for a wide range of parameters of the colliding objects. The results are succinctly described by a compact formula for the resulting encounter kernel, Eq.~\eqref{eqn:approximation}, which can be used to rapidly estimate encounter rates as a function of two dimensionless groups, the P\'eclet number $\peclet$ and $\beta$, the relative size of the objects and the large particle. In the context of marine snow, our work implies that advection-diffusion often remains the main driver of encounters with plankton and gels, even at very high P\'{e}clet numbers. Overall, by improving estimates of encounter rates, our results can inform models of carbon cycling in ocean ecosystems.

\section*{Data availability statement}

All the software used in the above simulations is open source with an open license and can be accessed from our repositories on Github \citep{gh_pypesh} and Zenodo \citep{Waszkiewicz_2025}. Additionally, as mentioned earlier, the Python package \code{pypesh} accompanies this manuscript. This package interpolates the $\sherwood$ values from the calculations presented in Figure~\ref{fig:sh_vs_pe} or performs direct computations for given parameters.

\section*{Acknowledgments}
The authors thank the Fenix Science Club for computational power. The authors thank Dr. Emilia Trudnowska for sharing images of marine particulate matter and insightful comments.
The work was supported by a Swiss NSF Ambizione grant no. PZ00P2\textunderscore202188 to JS; and the National Science Centre of Poland Sonata Bis grant no. 2023/50/E/ST3/00465 to ML.

\bibliography{sources}

\clearpage

\end{document}

\title{
Bridging advection and diffusion in the encounter dynamics of sedimenting marine snow \\
Supplementary Materials
}

\author{Jan Turczynowicz}
\thanks{These authors contributed equally to this work.}
\affiliation{Faculty of Physics, University of Warsaw, Pasteura 5, 02-093 Warsaw, Poland}
\affiliation{Fenix Science Club, Aleja Stanów Zjednoczonych 24, 03-964 Warsaw, Poland.}

\author{Radost Waszkiewicz}
\thanks{These authors contributed equally to this work.}
\affiliation{Institute of Physics, Polish Academy of Sciences, Aleja Lotnikow 32/46, PL-02668 Warsaw, Poland}
\affiliation{Fenix Science Club, Aleja Stanów Zjednoczonych 24, 03-964 Warsaw, Poland.}

\author{Jonasz Słomka}
\thanks{Corresponding authors: jslomka@ethz.ch (JS); mklis@fuw.edu.pl (ML).}
\affiliation{Institute of Environmental Engineering, Department of Civil, Environmental and Geomatic Engineering, ETH Zurich, Zurich, Switzerland}

\author{Maciej Lisicki}
\thanks{Corresponding authors: jslomka@ethz.ch (JS); mklis@fuw.edu.pl (ML).}
\affiliation{Faculty of Physics, University of Warsaw, Pasteura 5, 02-093 Warsaw, Poland}

{
\let\clearpage\relax
\maketitle
}

All of the plots presented in the main text are available in the GitHub repository, (see \url{https://github.com/turczyneq/pypesh}). Figures~1, 2, 5 and 6 are located under \code{examples/}, Figure~3 is under \code{literature\_comparison/} with data from the cited literature, and Figure~4 can be found in \code{results/}.

\section{Mesh tests}

\begin{figure}[h!]
    \centering
    \begin{tabular}{c}
         \includegraphics[width=0.6\linewidth,valign=t]{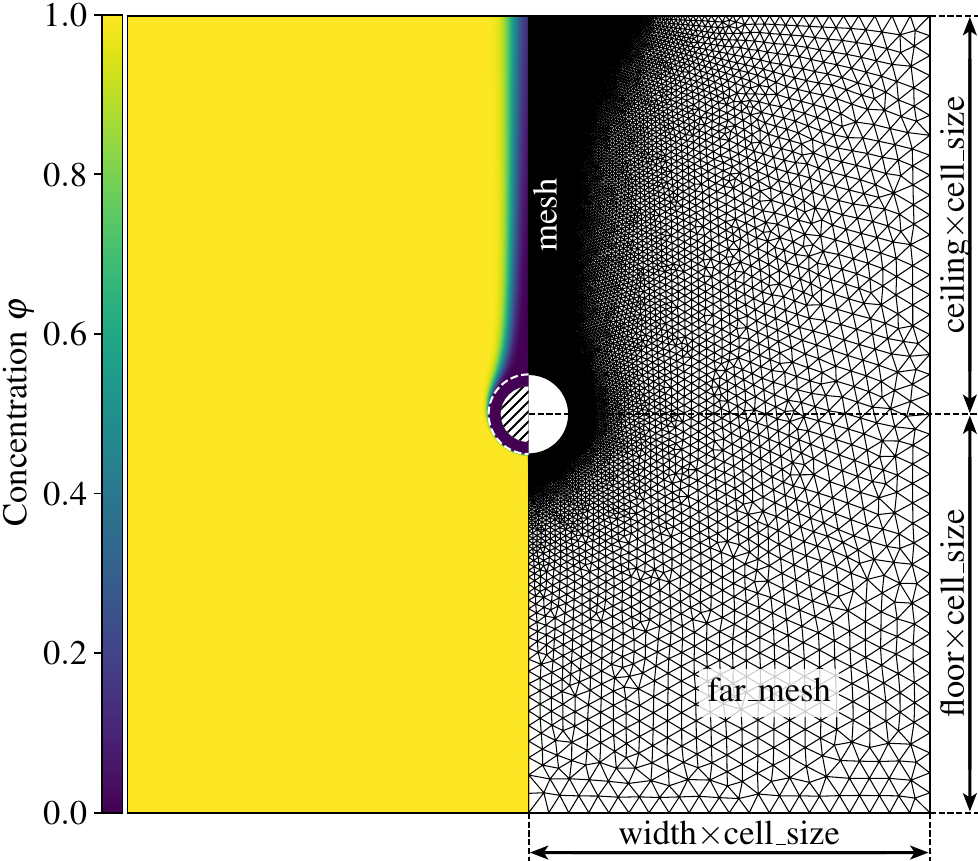}         
    \end{tabular}    
    \caption{
The figure shows an example of the solution (left) and mesh (right) for the finite element method approach. The marked parameters were selected to achieve convergence (see Figure~S2). On repository under \code{appendix/figure\_1.py}
    }
    \label{fig:mesh}
\end{figure}

We performed our finite element solution on a mesh specifically designed for our equation, as shown in Figure~\ref{fig:mesh}. We adjusted the \code{mesh} size in the most interesting region and modified the \code{far\_mesh} size away from this region. Additionally, we varied \code{width}, the distance from the sphere to the \code{ceiling} and \code{floor}. The size of the entire cell could also be adjusted using the parameter \code{cell\_size}.

\begin{figure*}[h!]
    \centering
    \begin{tabular}{c}
         \includegraphics[width=0.9\linewidth,valign=t]{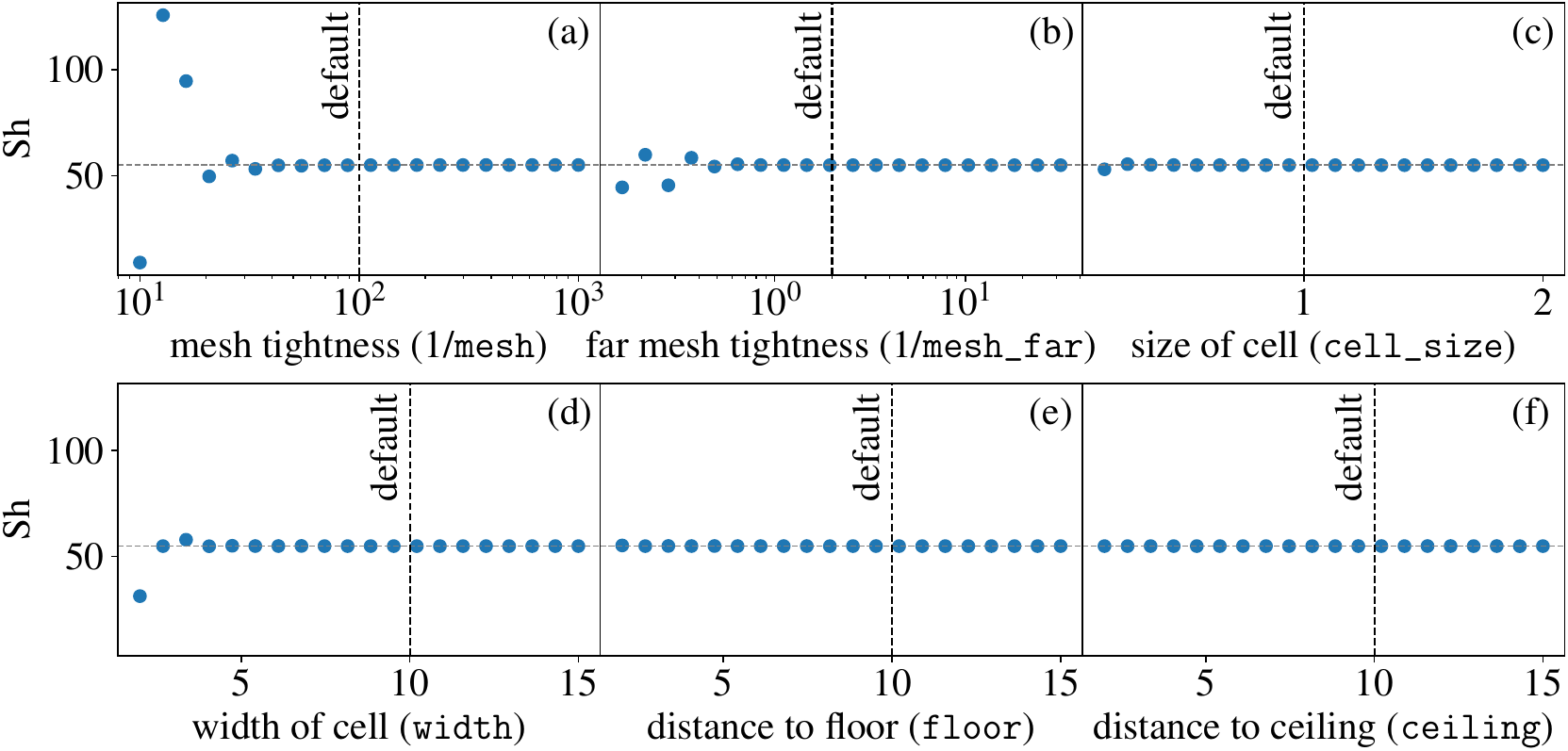}         
    \end{tabular}    
    \caption{
    Convergence tests for different values of numerical parameters (Figure~S1) that could affect our results. Vertical lines denote the default values used in our solution. The plot was generated for the following values of $\peclet=10^4$ and $\rparticle=0.1$. The plot is available under \code{appendix/mesh\_tests.py}.
    }
    \label{fig:mesh_test}
\end{figure*}

We studied the convergence of the method used by varying these values. A typical result of such a test is shown in Figure~\ref{fig:mesh_test}. In most of our calculations, $\code{mesh} = 0.01$, $\code{far\_mesh} = 0.5$, $\code{cell\_size} = 1$, and $\code{width} = \code{ceiling} = \code{floor} = 10$. Since this approach was significantly more efficient in terms of computation time, the default values were chosen rather conservatively.

\begin{figure}[h!]
    \centering
    \begin{tabular}{c}
         \includegraphics[width=0.9\linewidth,valign=t]{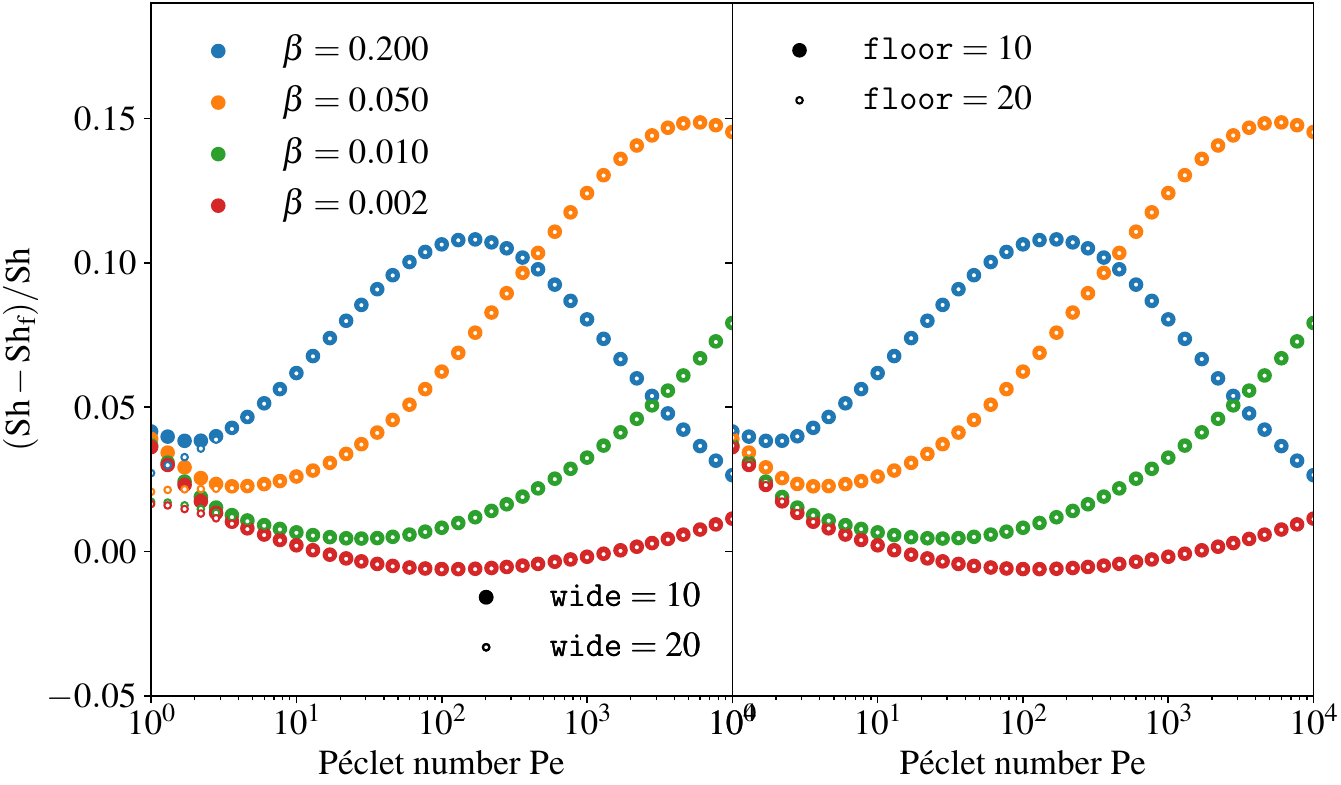}
    \end{tabular}    
    \caption{
    Convergence test for \code{width} and \code{floor} for different values of $\peclet$ and $\rparticle$. Plot available under \code{appendix/changing\_wide\_floor.py}.
    }
    \label{fig:mesh_test_pe_beta}
\end{figure}

Additionally, we compared two values of \code{floor} and \code{width} for different $\peclet$ and $\rparticle$, as shown in Figure~\ref{fig:mesh_test_pe_beta}.

\newpage

\section{Efficient sampling of hitting probability}

\begin{figure}
    \centering
    \begin{tabular}{c}
         \includegraphics[width=0.9\linewidth,valign=t]{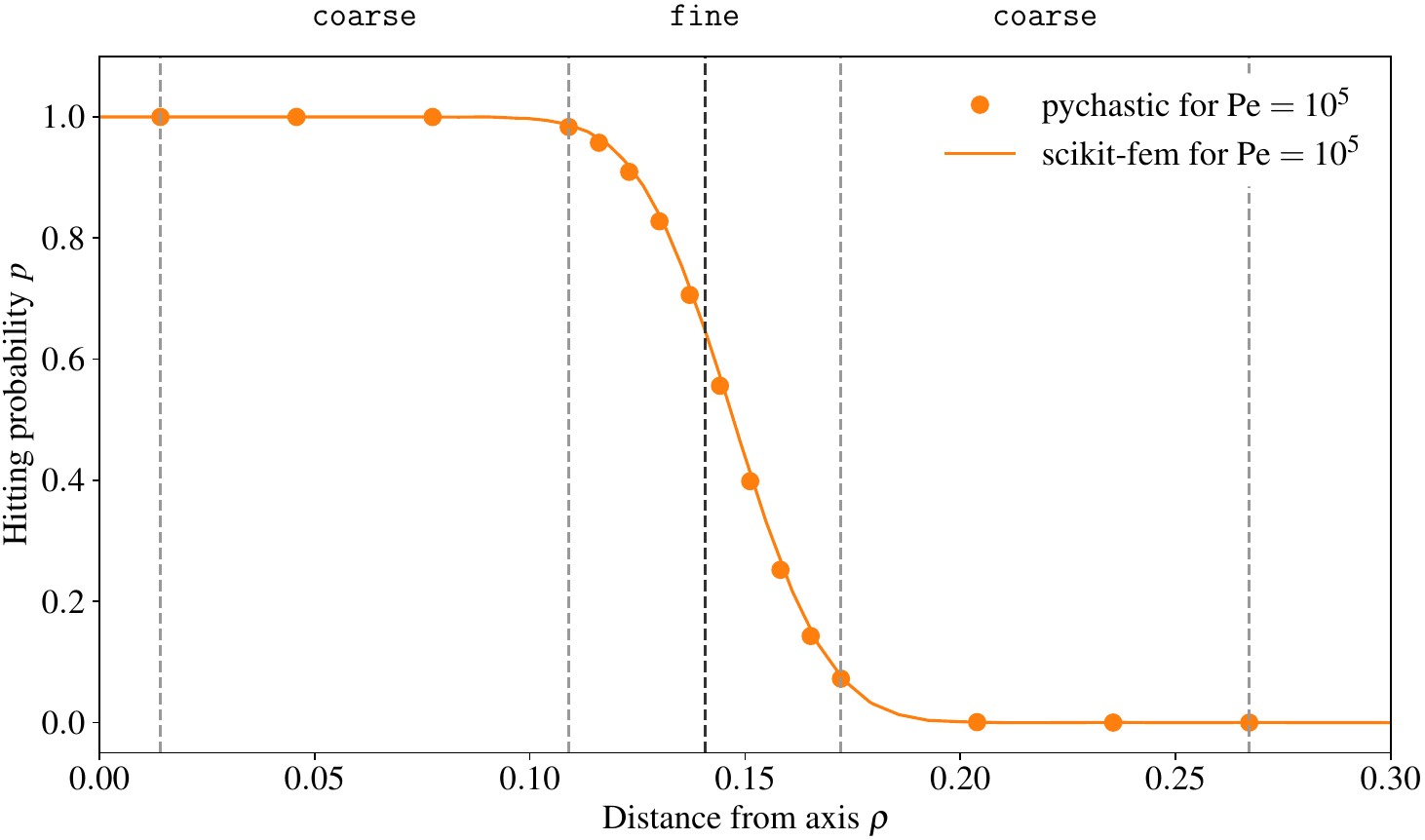}         
    \end{tabular}    
    \caption{
    Visualisation of the method used to calculate the flux onto the particle. The number of points and size of the coarse region was decreased to $40 \cdot (1/\peclet)^{1/2}$ for better visibility. Probability at each point is estimated based on $N=10^4$ stochastic trajectories. The plot is available under \code{appendix/spread\_explanation.py}.
    }
    \label{fig:dispersion}
\end{figure}

In the trajectories approach, determining the probability of hitting at a specific $x_0$ was computationally expensive, so an efficient distribution of sampling was required. The first step was to determine the radius of the streamline, $r_s$, which for $z = 0$ lies on $\rho = 1$. Based on observations, it was estimated that the dispersion of the step function at $r_s$ behaves as $(1/\peclet)^{1/2}$, as shown in Figure~\ref{fig:dispersion}, with a comparison to the distribution from \code{scikit-fem}.

\begin{figure}
    \centering
    \begin{tabular}{c}
         \includegraphics[width=0.9\linewidth,valign=t]{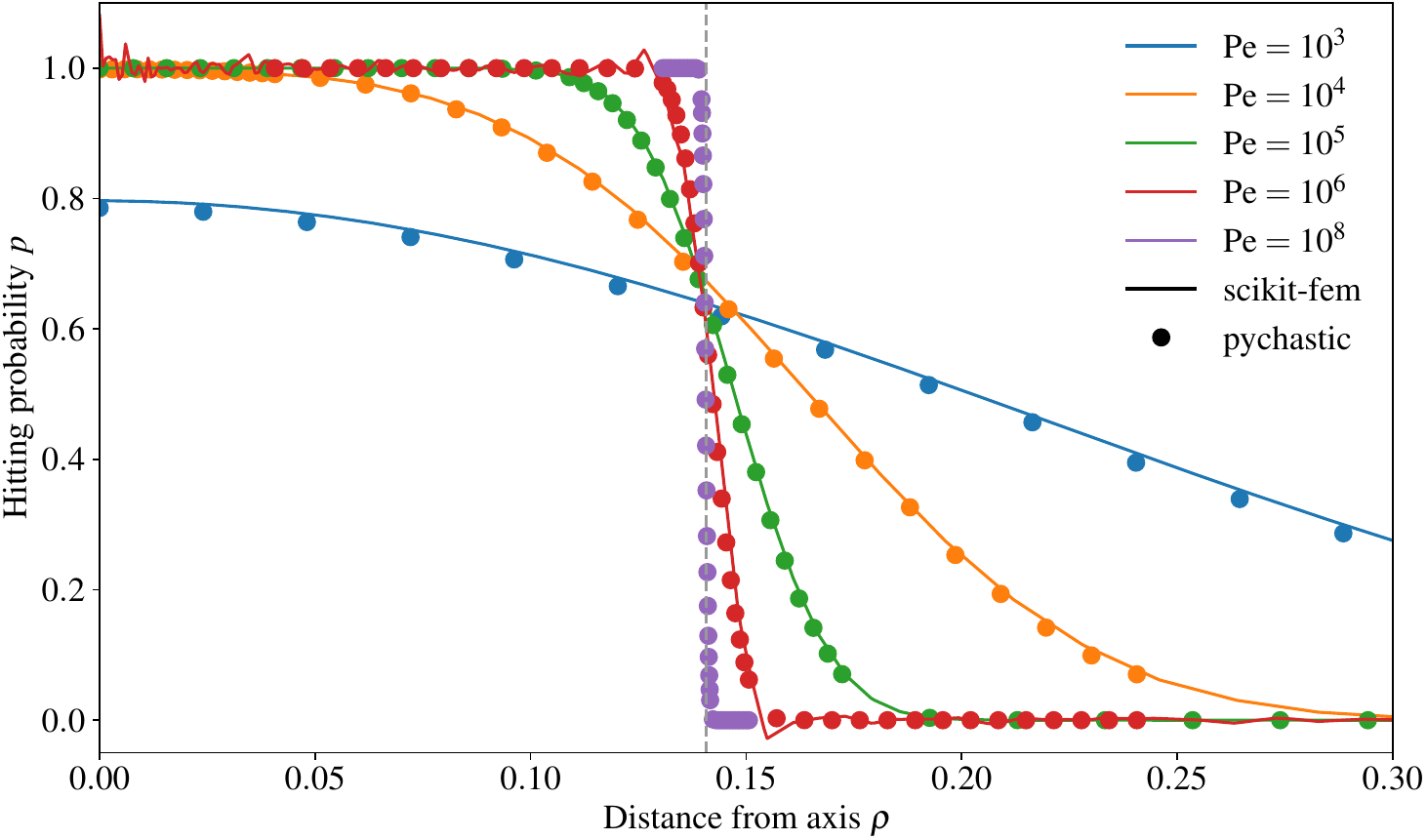}         
    \end{tabular}    
    \caption{
     The graph shows the cross-sections of the hitting probability generated via the finite element method (lines) and the stochastic trajectories approach (circles). For $\peclet=10^6$, the finite element method starts to show errors due to a high slope of the solution. In contrast, the method based on stochastic trajectories works very well, thus allowing for correct calculation of $\sherwood$. The probability distribution from the trajectories approach for $\peclet=10^8$ shows a good estimation of the spread region. The plot is available under \code{appendix/drawing\_cross\_sections.py}.
    }
    \label{fig:cross_sections}
\end{figure}

This allowed for the effective prediction of the region of greatest interest and significantly reduced the number of required points where the probability was calculated. The actual approach involved finding $r_s$ and setting test hitting probabilities densely (20 points) around it at a distance of $10 \cdot (1/\peclet)^{1/2}$, then adding more sparsely distributed test points (15 points on each side) at distances up to $100 \cdot (1/\peclet)^{1/2}$. Here, again, distances were chosen rather conservatively, as shown in Figure~\ref{fig:cross_sections}, where the coarse region ends at a constant value of 1 or 0.

\section{Tests of compatibility between methods}
\label{sec:compatibility}

The first comparison was performed by plotting cross-sections at a selected height, generated using both the finite element method and trajectories. The results can be seen in Figure~\ref{fig:cross_sections}. It is clear that the probability predictions are in agreement and that the predicted spread positions, based on our heuristic observations, are accurate, allowing us to greatly reduce computation time. For high $\peclet$, FEM shows signs of introducing errors and artifacts due to the proximity of very high slopes.

\begin{figure}
    \centering
    \begin{tabular}{c}
         \includegraphics[width=0.9\linewidth,valign=t]{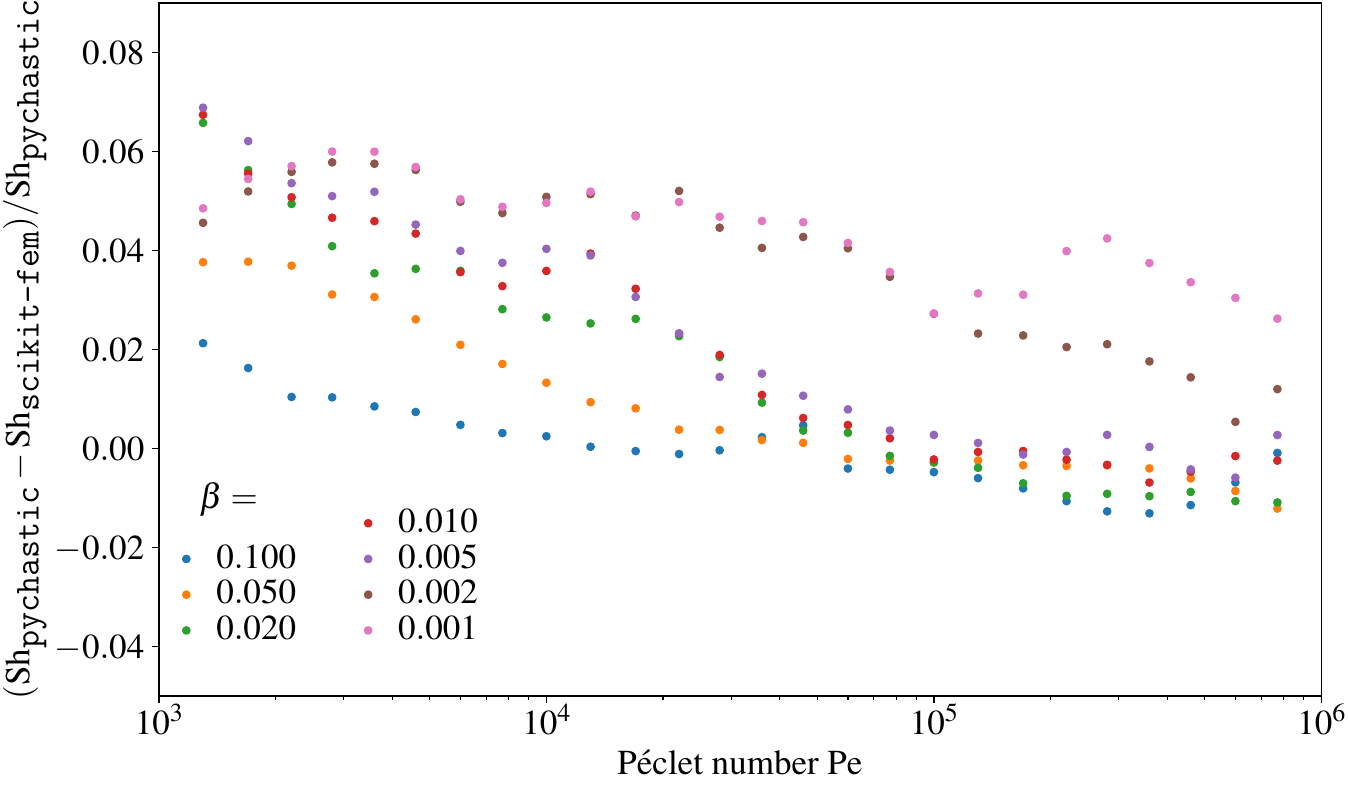}         
    \end{tabular}    
    \caption{
    Relative error between the two numerical methods for various values of $\peclet$ and $\rparticle$.
    }
    \label{fig:difference}
\end{figure}

A more accurate comparison was obtained by calculating the difference between the Sherwood numbers computed using the two methods for different $\peclet$. The results are presented in Figure~\ref{fig:difference}. It is clear that over a wide range of $\peclet$ and for all values of $\rparticle$, predictions from both methods are very close, with the relative difference remaining below 5\% for $\peclet>10^4$.

\section{Calculation parameters}

The run scripts are available in the GitHub repository under \code{results/run\_codes}. The value of the hitting probability was estimated based on $N = 10^4$ trials. In the coarse regions, there were 15 points each, while in the fine region, there were 20 points.

For higher values of $\peclet$, the probability of a tracer being pushed out of the streamline close to the particle was low. As the flow velocity near the particle decreases significantly, the time for these tracers to travel around the particle increases. Thus, for small $\rparticle$ and high $\peclet$, it was necessary to adjust the simulation time. For $\beta > 0.005$ and $\beta = 0.005$ with $\peclet \leq 10^6$, the simulation time \code{t\_max} was set to 200; for $\beta = 0.005$ with $\peclet > 10^6$, $\beta = 0.002$, and $\beta = 0.001$ with $\peclet < 10^5$, \code{t\_max} was set to 800; and for $\beta = 0.001$ with $\peclet \geq 10^5$, \code{t\_max} was set to 1600.

\newpage

\section{Discussion of Figure 6}

\begin{figure}
    \centering
    \begin{tabular}{c}
         \includegraphics[width=0.9\linewidth,valign=t]{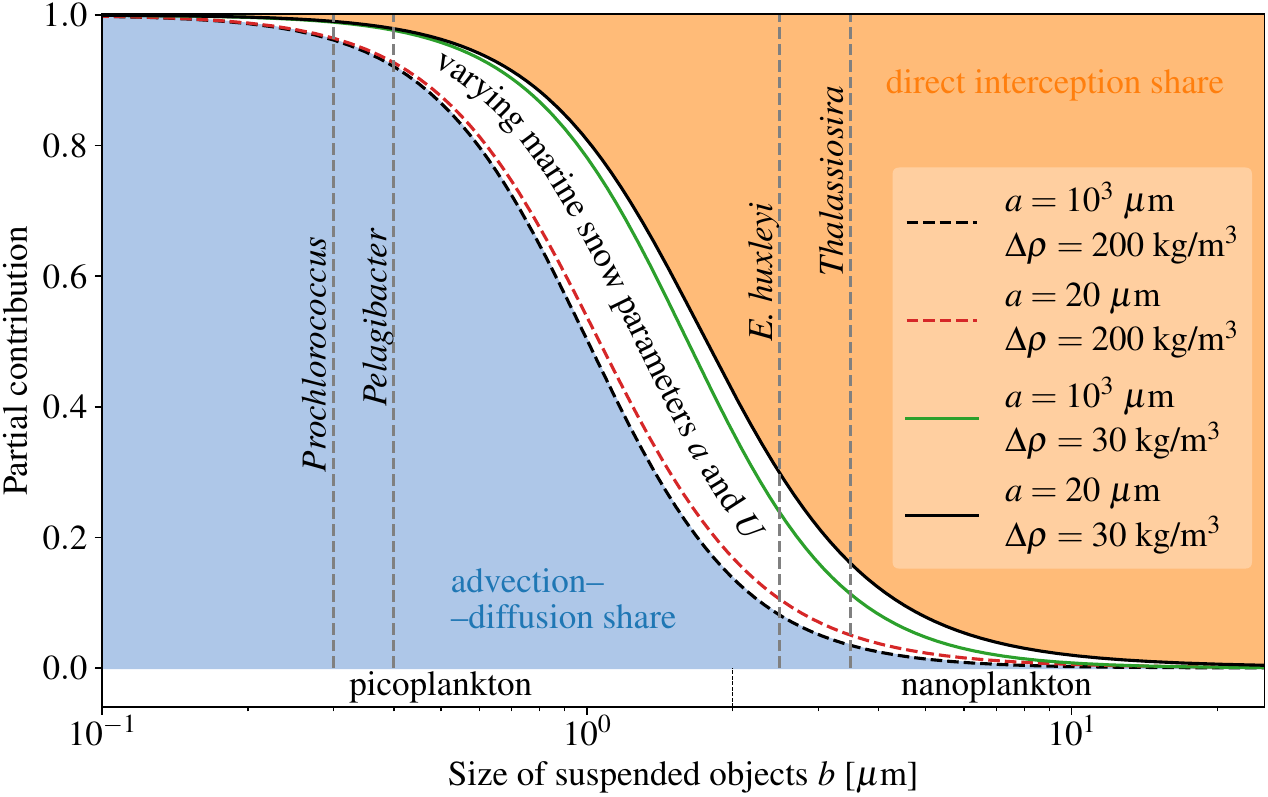}         
    \end{tabular}    
    \caption{
    Investigation of the influence of the sinking particle size on the white region in Figure~6 of the main text. The plot is the same as the plot shown in Figure 6 with additional lines marking the changes induced by varying the particle size. Specifically, for the boundary densities $\Delta \rho=30\;\textrm{kg}/\textrm{m}^3$ and $\Delta \rho=200\;\textrm{kg}/\textrm{m}^3$, both boundary particle sizes are plotted (green and red lines). Despite the change of $a$ for two orders of magnitude, the partial contribution of each encounter mechanism changes only slightly. 
    }
    \label{fig:a_dont_matter}
\end{figure}

Figure~\ref{fig:a_dont_matter} shows how a change in particle size $a$ affects the partial contribution. Despite a change of two orders of magnitude, the shares of $\Phi_\textrm{A}$ and $\Phi_\textrm{Cl}$ remain similar, indicating that the type of collision weakly depends on the particle size $a$ (assuming the sinking speed follows Stokes' law).

\begin{figure}
    \centering
    \begin{tabular}{c}
         \includegraphics[width=0.9\linewidth,valign=t]{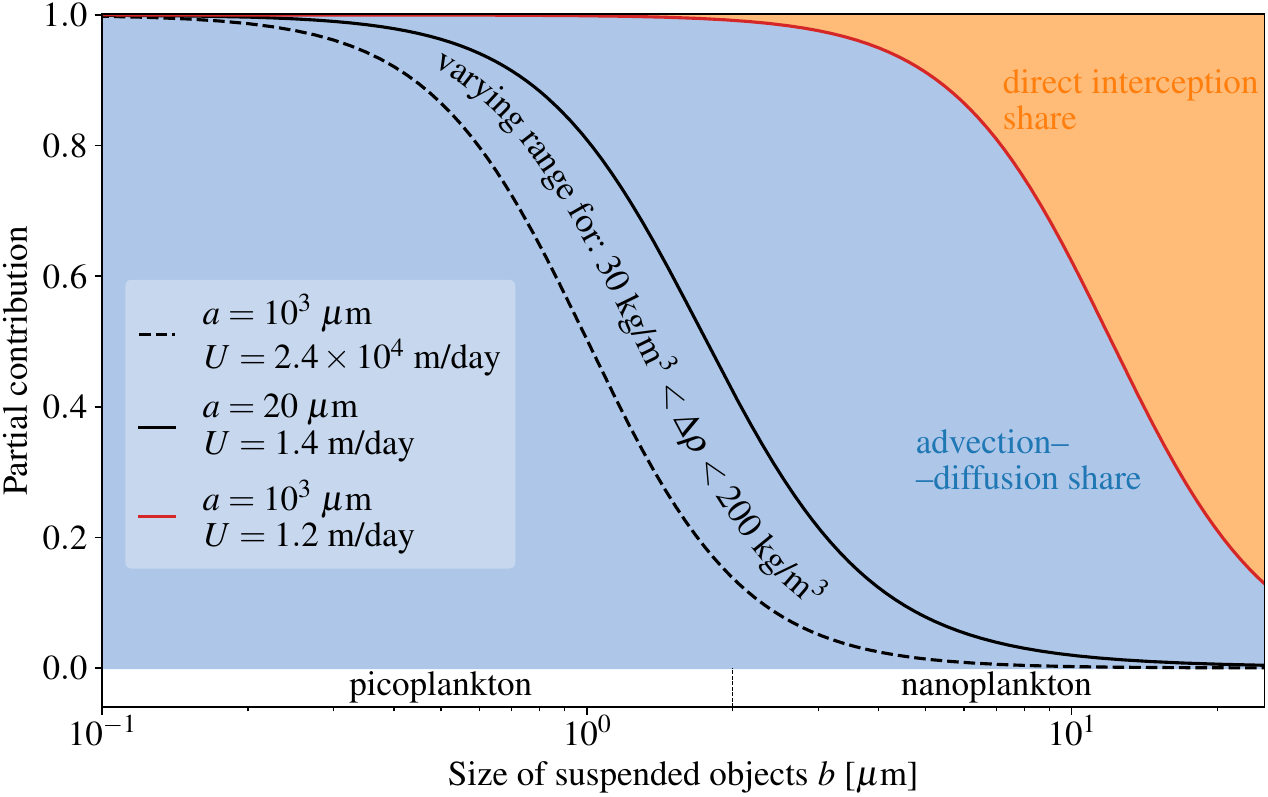}         
    \end{tabular}    
    \caption{
    A variant of Fig. 6 in the main text for the case of sinking particles with very small density difference $\Delta \rho=0.01\;\textrm{kg}/\textrm{m}^3$. For comparison, the black lines mark the density difference range assumed in Figure~6. For such slowly sinking marine snow particles, advection-diffusion remains an important mechanism generating encounters with objects as large as $b=10~\mu\textrm{m}$ (red line).
    }
    \label{fig:low_rho}
\end{figure}

On the other hand, changes in $ \Delta \rho $, specifically when the density difference is smaller than the assumed boundary $ \Delta \rho = 30\;\textrm{kg}/\textrm{m}^3 $, affect the white region.  
For slowly sinking particles, one can consider a particle with $ \Delta \rho \approx 10^{-2}\;\textrm{kg}/\textrm{m}^3 $, which corresponds to $ U = 1.2\;\textrm{m}/\textrm{day} $ for $ a = 10^3\;\mu\textrm{m} $. Figure~\ref{fig:low_rho} shows that the region dominated by advection-diffusion collisions shifts toward larger $ b $. This implies that as density decreases, an increasing number of collisions are driven by advection-diffusion, leading to greater underestimation compared to the case when the direct interception collision mechanism alone is assumed.

